\documentclass[10pt]{article}

\usepackage{graphicx}%
\usepackage{multirow}%
\usepackage[nosumlimits,nonamelimits]{amsmath}%
\usepackage{amssymb,amsfonts}%
\usepackage{amsthm}%
\usepackage{mathrsfs}%
\usepackage[title]{appendix}%
\usepackage{xcolor}%
\usepackage{textcomp}%
\usepackage{manyfoot}%
\usepackage{listings}%
\usepackage[sort]{natbib}
\numberwithin{equation}{section}

\newtheorem{theorem}{Theorem}
\newtheorem{assumption}{Assumption}
\newtheorem{remark}{Remark}
\newtheorem{algorithm}{Algorithm}

\usepackage[sort&compress]{cleveref}
    \crefname{axiom}{Axiom}{Axioms}
    \crefname{claim}{Claim}{Claims}
    \crefname{corollary}{Corollary}{Corollaries}
    \crefname{hypothesis}{Hypothesis}{Hypotheses}
    \crefname{lemma}{Lemma}{Lemmata}
    \crefname{proposition}{Proposition}{Propositions}
    \crefname{theorem}{Theorem}{Theorems}
    \crefname{algorithm}{Algorithm}{Algorithms}
    \crefname{assumption}{Assumption}{Assumptions}
    \crefname{definition}{Definition}{Definitions}
    \crefname{example}{Example}{Examples}
    \crefname{fact}{Fact}{Facts}
    \crefname{notation}{Notation}{Notations}
    \crefname{property}{Property}{Properties}
    \crefname{remark}{Remark}{Remarks}
    \crefname{equation}{Eq.}{Eqs.}
    \crefname{figure}{\textsc{Fig}.}{\textsc{Fig}s.}
    \crefname{table}{\textsc{Tab.}}{\textsc{Tab}s.}

\DeclareMathOperator*{\argmax}{arg\,max}

\DeclareMathOperator*{\argsup}{arg\,sup}

\usepackage{booktabs}
\usepackage{tabularx}
    \newcolumntype{L}{>{\raggedright\arraybackslash}X}
    \newcolumntype{C}{>{\centering\arraybackslash}X}
    \newcolumntype{R}{>{\raggedleft\arraybackslash}X}

 \usepackage{enumitem}
    \newlist{propenum}{enumerate}{1} 
    \setlist[propenum]{label={(\alph*.)},ref=\theassumption\alph*}
    \crefalias{propenumi}{assumption}

\usepackage{tikz}
    \usetikzlibrary{calc}
    \usetikzlibrary{positioning}
    
\usepackage{algpseudocode}
    \algrenewcommand\algorithmicdo{\textbf{\textsc{do}}}
    \algrenewcommand\algorithmicelse{\textbf{\textsc{else}}}
    \algrenewcommand\algorithmicend{\textbf{\textsc{end}}}
    \algrenewcommand\algorithmicensure{\textbf{\textsc{Return:}}}
    \algrenewcommand\algorithmicfor{\textbf{\textsc{for}}}
    \algrenewcommand\algorithmicif{\textbf{\textsc{if}}}
    \algrenewcommand\algorithmicrequire{\textbf{\textsc{Require:}}}
    \algrenewcommand\algorithmicthen{\textbf{\textsc{then}}}
    \algrenewcommand\algorithmicwhile{\textbf{\textsc{while}}}

\usepackage{mathtools}

\title{\vspace{-3cm} Bandit Algorithms for Policy Learning: Methods, Implementation, and Welfare-performance}

\date{\today}
\author{
        Toru Kitagawa\thanks{Department of Economics, Brown University. Email: toru\_kitagawa@brown.edu}
            \and
        Jeff Rowley\thanks{Department of Economics, University College London.}
}

\begin{document}
\maketitle

\vspace{-1cm} 

\abstract{%
Static supervised learning---in which experimental data serves as a training sample for the estimation of an optimal treatment assignment policy---is a commonly assumed framework of policy learning. 
An arguably more realistic but challenging scenario is a dynamic setting in which the planner performs experimentation and exploitation simultaneously with subjects that arrive sequentially.
This paper studies bandit algorithms for learning an optimal individualised treatment assignment policy. Specifically, we study applicability of the EXP4.P (Exponential weighting for Exploration and Exploitation with Experts) algorithm developed by \citet{EXP4.P} to policy learning.
Assuming that the class of policies has a finite Vapnik-Chervonenkis dimension and that the number of subjects to be allocated is known, we present a high probability welfare-regret bound of the algorithm.
To implement the algorithm, we use an incremental enumeration algorithm for hyperplane arrangements. 
We perform extensive numerical analysis to assess the algorithm's sensitivity to its tuning parameters and its welfare-regret performance. 
Further simulation exercises are calibrated to the National Job Training Partnership Act (JTPA) Study sample to determine how the algorithm performs when applied to economic data. 
Our findings highlight various computational challenges and suggest that the limited welfare gain from the algorithm is due to substantial heterogeneity in causal effects in the JTPA data.}


\textbf{Acknowledgements}: Toru Kitagawa is honoured to accept the 2023 Nakahara prize by the Japanese Economic Association, and thanks the selection committee for their consideration. The authors thank the editor and a referee for beneficial comments, and gratefully acknowledge financial support from ERC grant 715940,
the ESRC Centre for Microdata Methods and Practice (CeMMAP) (RES-589-28-0001).

\newpage

\section{Introduction}\label{SEC:INTRODUCTION}

How to make use of evidence for policymaking is a topic of great importance.
The growing literature on statistical treatment choice \citep{manski2004statistical, dehejia2005program} and on policy learning \citep[etc.]{KT, KT21, mbakop2021model, athey2021policy, kitagawa2021constrained} develops formal frameworks and methods for combining causal inference with the social planner's decision, making use of statistical decision theory and machine learning methods. 
A commonly assumed setting is that of static supervised learning, wherein experimental data serves as training data and learning an optimal treatment assignment policy happens only once. 
The attraction of this setting is its simplicity;
it ignores, however, the important dynamic aspects of learning and of implementation of treatment assignment policies. 
Subjects that are assigned to treatments and that contribute to causal evidence often appear sequentially over time. 
Accordingly, operations to accumulate evidence, learn causal effects, and assign treatments can run simultaneously over multiple time periods. 
In such a dynamic setting:
Which strategy should the social planner implement to maximise a welfare criterion? 
Can we expect this strategy to deliver substantial welfare gains in public policy applications? 
How does this strategy compare with the static supervised learning approach, in terms of the assignment policy that it implements and the welfare that it generates? 
And, is it feasible to compute and implement a sophisticated dynamic sampling and assignment strategy in practice?

This paper aims to answer these questions by formulating dynamic policy learning as a multi-arm bandit problem with contextual information. 
Multi-arm bandit problems capture dynamic environments in which a new subject arrives in every period, with the decision-maker assigning the subject to one of the many candidate treatments that are available. 
The decision-maker learns the effects of these treatments from the subjects' responses, and updates the rules by which subjects are assigned to treatments sequentially.
The goal of the analysis is to find rules that maximize the sum of the subjects' treatment outcomes by effectively balancing exploration and exploitation, for which there is a trade-off.
In the bandit problem setting, contextual information refers to the observable pre-treatment characteristics of the sequentially arriving subjects, upon which the decision-maker can discriminate (i.e., by ascribing different treatment rules to subjects with different pre-treatment characteristics). 
Individualised assignment based upon contextual information outperforms non-individualised assignment if treatment effects are heterogeneous across the pre-treatment covariates. 

Our approach is to associate an arm in a bandit problem with an individualised treatment assignment rule that maps a subject's pre-treatment covariates to an assigned treatment. 
We then propose a bandit algorithm to learn an optimal individualised treatment assignment policy from a class of policies. 
We assume that the class of individualised treatment assignment policies is finite or is infinite with finite Vapnik-Chervonenkis (VC) dimension, as is considered in \citet{KT}. 
Hence, we deal with the challenging situation of a bandit problem in which there is an infinite number of arms---something that arises if a subject's pre-treatment covariates include a continuous variable.
In contrast to contextual linear bandit problems---like those studied in \cite{Auer_2002} and \cite{Abbasi_Yadkori_et_al_2011}, to list but a few relevant papers---our approach does not impose functional form restrictions on the conditional average treatment effect.

Our approach builds upon the influential work of \citet{Auer_etal_2002} that proposes the EXP4  (Exponential weighting for Exploration and Exploitation with Experts) algorithm for an adversarial bandit. 
Extending the original EXP4 algorithm, \citet{EXP4.P} proposes the EXP4.P algorithm for learning classification rules with a finite VC dimension---which we denote by $D < \infty$. 
Both algorithms consider a set of experts and probability distributions over them.
Each expert gives a recommendation (i.e., which bandit arm to pull given contextual information). 
The decision-maker makes their decision by aggregating the experts' recommendations according to a probability distribution. 
Every time a decision is made, the decision-maker receives feedback on each expert's performance, updating the probability distribution over the experts that they use based upon this feedback. 
Viewing a assignment rule as an expert, \citet{EXP4.P} shows that iterative implementation over a number of periods---which we denote by $T < \infty$---attains a high-probability upper bound on the distribution of cumulative regret of order $\mathrm{O}( \sqrt{T \cdot D \cdot \ln (T)} )$, implying that the average welfare-regret converges at a rate of $\mathrm{O}( \sqrt{D / T \cdot \ln (T) })$. 

We modify the EXP4.P algorithm to develop a version of the algorithm that is tailored to policy learning. 
Specifically, we associate the aforementioned set of experts with a class of individualised treatment assignment rules that is finite or of finite VC dimension.
Our modified algorithm accommodates treatment choice given an additive welfare criterion and a general bounded outcome variable, and we show that, for suitable choices of tuning parameters, it achieves a high-probability upper bound on the cumulative welfare-regret of order $\mathrm{O}( \sqrt{T \cdot D \cdot \ln (T)} )$.  

An important step in implementing the EXP4.P algorithms to coarsen the class of assignment rules to a finite class using information gathered from the first $\tau$ observations. 
This step is computationally non-trivial and, to our knowledge, development of computationally efficient methods for this step is an open question. 
Focusing on the class of Linear Eligibility Score (LES) rules, we show that the coarsening problem is equivalent to a hyperplane arrangement problem in the geometry literature.
Various cell enumeration procedures have been proposed in that literature, and we suggest using the Incremental Enumeration algorithm introduced by \citet{RS} \citep[and improved upon by][]{GU} as one way to conduct this step.

To illustrate implementation of our modified EXP4.P algorithm and to assess its welfare-performance, we perform extensive simulation studies. 
In one simulation design, we modify the variances of the potential outcomes (i.e., the heterogeneity of individual causal effects) whilst holding fixed the magnitude of conditional average treatment effects. 
We find that the (median and variance of the) welfare-performance of the EXP4.P algorithm is sensitive to the heterogeneity of individual causal effects, indicating a challenge in those policy applications where subjects' treatment responses are highly heterogeneous in unobservables. 
We also assess the sensitivity of this welfare-performance to the tuning parameters that govern the exploration and exploitation trade-off of the algorithm. 

Through a novel simulation design that we calibrate to actual economic data, we also investigate how our modified EXP4.P algorithm performs in public policy applications.
We use data from the National Job Training Partnership Act (JTPA) Study to do this. 
The original JTPA sample contains the time stamp of when each experimental subject entered the Study. 
Maintaining the order of individual entry, we perform a counterfactual analysis of the welfare-level that the social planner would attain if assignment were based upon the EXP4.P algorithm. 
We estimate potential outcome regressions via random forest methods using the JTPA sample, and construct distributions of potential outcomes from these and their residuals.
By applying our modified EXP4.P algorithm to a multitude of samples drawn from these distributions, we obtain a distribution over the average welfare that the algorithm attains.

Recent research in policy learning extends to and intersects with the machine learning literature on bandit algorithms \citep[see][for a monograph on bandit algorithms]{Lattimore_2020}.
The welfare-performance criterion that we consider concerns the cumulative welfare for sequentially arriving units rather than for the super-population. The latter corresponds to the welfare objective in the best-arm identification problems as studied in \citep{Russo_etal_2016,kasy_2021,Ariu_2021}, among others. \citet{Athey_et_al_2022} and \citet{Qin_et_al_2024} assess the trade-off between the in-sample welfare and the super-population welfare of best-arm identification and study how to balance them out. 
Recent advances in bandit algorithms in the econometrics literature include those studied in \citet{Adusumilli_2021,kock_2022,Dimakopoulou_et_al_2017,Kuang_2023}, to list but a few relevant papers. 
Application and feasible implementation of EXP4.P algorithms to policy learning have, to our knowledge, not been studied in the policy learning literature. 

Preceding the EXP4.P algorithm, the literature of online learning has studied exponential weighting of experts in prediction and portfolio choice problems; see \citet{cesa2006prediction}, \citet{Littlestone_et_al_1994}, and \citet{Vovk_1990} for a monograph and early works of the topic. \cite{Chen_2023} and \cite{Viviano_et_al_2023} apply the perspective and methods of online aggregation of experts to causal inference with panel data.  

This paper's dynamic approach to policy learning is different from the approach considered in the literature on dynamic treatment regimes \citep{Robins_1986,Murphy_2003, Zhang_et_al_2018,Ko_2022,ida2024dynamic, Han_2023}.
That literature considers estimation of adaptive allocation of treatments to the same individual over multiple time periods using exogenous training data. 
For a similar reason, this paper differs from the time-series Empirical Welfare Maximisation (EWM)-approach proposed in \citet{kitagawa2024timeseries}. 
Exponential weighting over classification rules or individualised treatment rules based upon their empirical performance appears, however, in Probably Approximately Correct (PAC) Bayes analysis present in the supervised learning literature \citep[see][and references therein]{McAllester2003,Begin2014,kitagawa2022stochastic}.

\section{Framework}
A utilitarian social planner is faced with a sequential allocation problem of $K$ treatment arms over $T$ periods---both of which are finite and are known to the social planner.
The allocation problem involves the repeated interaction of the social planner with a stable and passive environment (Nature and subjects).  
The social planner observes subjects' responses to realised treatment arms and bases her choice of treatment in subsequent periods, in part, upon these responses.

\subsection{Timing and the flow of information}
\label{SEC:TIMING}
A subject---here, subject $t$---is characterised by a collection of $J$ features together with a collection of $K$ counterfactual responses (i.e., by the subject's response to each treatment arm). 
We denote the collection of features by $\mathbf{x}_{t} \in \mathcal{X} \subset \mathbb{R}^{J}$ and refer to these as covariates, with $x_{t,j}\in\mathbb{R}$ denoting the $j$th covariate.
We denote the collection of counterfactual responses by $\mathbf{y}_{t} = (y_t(1), \dots, y_{t}(K))^{\intercal} \in \mathbb{R}^{K}$ and refer to these as potential outcomes, with $y_{t}(k) \in \mathbb{R}$ denoting the potential outcome corresponding to the $k$th treatment arm.
We denote the population by $P$, which constitutes a joint distribution over covariates and potential outcomes that we concatenate as $(\mathbf{y}_{t}^{\intercal}, \mathbf{x}_{t}^{\intercal})$. 
We maintain the following assumptions on the population throughout.

\begin{assumption}[Stationary population]
\label{ASS:STATIONARY}
A sequence of random variables $(\mathbf{y}_{t}^{\intercal},\mathbf{x}_{t}^{\intercal})$ is independently and identically distributed to $P$. 
\end{assumption}

\begin{remark}
\Cref{ASS:STATIONARY} is standard in stochastic bandit problems.
We impose this assumption in our numerical analysis and so make this assumption here too.
A performance guarantee for welfare-regret under a nonstationary or adversarial environment is, however, known for EXP-type algorithms \citep[see][]{Lattimore_2020}.     
\end{remark}

\begin{assumption}[Bounded outcomes]
\label{ASS:UNIFORM}
$\mathbf{y}_{t}$ is such that $0\leq y_{t}(k) \leq M$ for all $k = 1, \dots, K$ and $t=1, \dots, T$.
\end{assumption}

\begin{remark}
\cref{ASS:UNIFORM} requires that outcome data is both bounded and non-negative.
Whilst economic data can always be bounded, but may not be guaranteed to be non-negative. 
Via the addition of a positive constant, economic data can be made non-negative to satisfy \cref{ASS:UNIFORM}. 
As is discussed in \citet{KT} in the context of static policy learning, an estimated optimal policy is not generally invariant to the addition of constants to outcomes. 
See \cref{REMARK:INVARIANCE} below for further discussion.    
\end{remark}


At the beginning of each period, a single subject is randomly selected from the population.
We adopt the convention of labelling subjects according to the order in which they are selected (i.e., Nature selects subject $t$ in period $t$).
Once a subject is selected, their covariates are revealed to the social planner (i.e., the social planner observes $\mathbf{x}_{t}$).
The social planner then---directly or as the outcome of some randomisation---administers a treatment, which we denote by $k_{t}\in\{1,...,K\}$, and we assume that the subject fully complies with the assigned treatment.
The social planner then observes the realised outcome (i.e., the social planner observes $y_{t}(k_{t})$) without delay. On the other hand, the social planner does not observe the potential outcomes corresponding to unrealised treatment arms.
At the end of each period, the social planner retains their accumulated knowledge up to that point (i.e., the social planner adds $(\mathbf{x}_{t},k_{t},y_{t}(k_{t}))$ to the information that she possesses at the beginning of period $t$) and carries this through to the next period.
Nature continues to select subjects until the population is exhausted---we reiterate that $T$, in addition to $K$, is finite and is known.

\subsection{The social planner's actions}
We denote the information set of the social planner at the beginning of period $t$ by $\mathcal{I}_{t}$.
The information set of the social planner is empty at the beginning of the first period, and otherwise evolves according to
\begin{equation}
\mathcal{I}_{t+1}
=
(\mathcal{I}_{t},k_{t},\mathbf{x}_{t}^{\intercal},y_{t}(k_{t})),
\end{equation}
as per \Cref{SEC:TIMING}.
We emphasise that a subject's covariates are revealed to the social planner in advance of her choice (i.e., the information set of the social planner in period $t$ at the time that she makes her choice comprises $\mathcal{I}_{t}$ and $\mathbf{x}_{t}$).

Following the terminology of the online learning literature \citep[e.g.,][]{cesa2006prediction}, we define an \textit{expert} as $f: \mathcal{X} \to \Delta^{K}$---a mapping from covariates to a probability distribution over the set of treatments. 
Each expert specifies a time-invariant individualised assignment rule (as a function of covariates) that allows for randomised allocation. 
That is, $f(\mathbf{x}_t)$ constitutes a $K$-vector whose $k$th entry---which we denote by $f_{k}(\mathbf{x}_{t})$---specifies the probability that subject $t$ is assigned to treatment $k$.
For instance, when there are only two treatment arms, the single index policy $f(\mathbf{x}_t) = (1((1,\mathbf{x}_t^{\intercal}) \beta < 0), 1((1,\mathbf{x}_t^{\intercal}) \beta \geq 0))^{\intercal}$ that is often considered in the context of static treatment choice \citep[e.g.,][]{KT} corresponds to an expert that allocates one of two treatment arms deterministically according to a fixed threshold rule for the sign of $(1,\mathbf{x}_t^{\intercal}) \beta$.
We denote the set of experts by $\mathfrak{F}$. 

The social planner's goal is to efficiently learn a best expert (i.e., individualised treatment assignment) in $\mathfrak{F}$ from sequentially arriving subjects. 
We define a \textit{treatment assignment policy} (of the social planner) as a dynamic strategy for assigning subjects to treatment arms that learns the best-performing experts in $\mathfrak{F}$ using information contained in $\mathcal{I}_t$. 
Specifically, a treatment assignment policy (constructed upon available data) is given by a sequence $\{ \mathbf{p}_t : \mathcal{I}_t \times \mathcal{X} \to \Delta^{K}|t=1, \dots, T \}$, where $\mathbf{p}_{t}$ is an $\mathcal{I}_{t}$-measurable map that maps subject $t$'s covariates $\mathbf{x}_t$ to a probability for each treatment arm. 
As such, the treatment of subject $t$ is allocated according to $k_t  | (\mathcal{I}_{t}, \mathbf{x}_{t})  \sim \mathbf{p}_t(\mathbf{x}_t)$.

The learning algorithms that we consider in this paper build an assignment policy by aggregating experts in $\mathfrak{F}$ according to their performance up to period $t$. 
As we present in the next section, what drives the aggregation formula for $\mathbf{p}_t$ in the EXP algorithm and derivative algorithms is a probability distribution over $\mathfrak{F}$ that we denote by $\mathbf{q}_{t}$. 
It is convenient to separate the construction of $\mathbf{q}_{t}$ according to the cardinality of $\mathfrak{F}$.

\begin{assumption}[Complexity]
\label{ASS:COMPLEXITY}
Either:
\begin{propenum}
\item \label{ASS:COMPLEXITY-FINITE}
    Finite experts---$K<\infty$ and $\mathfrak{F}$ has finite cardinality equal to $N$; or
\item \label{ASS:COMPLEXITY-VC}%
    Experts with complexity controlled by a finite VC dimension---$K=2$ and $\mathfrak{F}$ consists of deterministic rules of the form $f(\mathbf{x}) = 1(\mathbf{x} \in G)$, $G \subset \mathcal{X}$, and
    the class $\mathcal{G} =\{ G \}$ that spans $\mathfrak{F}$ has finite VC dimension equal to $D$.\footnote{%
    We let $\mathbf{X}^{\ell} \doteq \{\mathbf{x}^1,\dots,\mathbf{x}^{\ell} \}$ be a non-empty finite set with $\ell$ points in $\mathcal{X}$. 
    Given $\mathcal{G}$---a class of subsets in $\mathcal{X}$---we define $N(\mathbf{X}^{\ell})=|\{ \mathbf{X}^{\ell} \cap G : G\in \mathcal{G} \}|$ as the number of different subsets of $\mathbf{X}^{\ell}$ picked out by $G \in \mathcal{G}$.
    The VC dimension of $\mathcal{G}$ is the largest $\ell$ such that $\sup_{\mathbf{X}^{\ell}} N(\mathbf{X}^{\ell}) = 2^{\ell}$ holds. 
    See \citet{Vapnik1998Book} for more detailed explanation and numerous examples.}
\end{propenum}
\end{assumption}

In \Cref{subsec:F-EXP4.P}, we study the case of \cref{ASS:COMPLEXITY-FINITE}, where the number of treatment arms can be more than two and some experts in $\mathfrak{F}$ are allowed to be probabilistic but, crucially, the number of experts is finite. 
In \Cref{subsec:VC-EXP4}, we study the case of \cref{ASS:COMPLEXITY-VC}, where the number of treatment arms is allowed to be infinite but, crucially, there are only two treatment arms---further extension to more than two treatment arms is beyond the scope of this paper---and the set of experts comprises deterministic rules whose complexity is controlled by a finite VC dimension. 
We present the EXP4.P algorithm for each case, respectively. 

\subsection{The social planner's preferences and regret}
The objective of the social planner is to maximise some welfare criterion. 
Following the literature on bandit algorithms, we assume that the objective function of the social planner is the sum of the subjects' outcomes, and that the social planner is risk-neutral.   

We define the empirical and average welfare attained by implementing the recommendations of a single expert $f$ throughout the entirety of the time horizon by
\begin{align}
\label{EQ:W_EMP}
\hat{W}_T(f)
& \doteq  \textstyle\sum_{t=1}^{T} f(\mathbf{x}_t)^{\intercal}\mathbf{y}_{t}, \\
\label{EQ:W}
W_T(f)
& \doteq  \mathrm{E}_P ( \textstyle\sum_{t=1}^{T} f(\mathbf{x}_t)^{\intercal}\mathbf{y}_{t})
= T \cdot \mathrm{E}_{P}( f(\mathbf{x}_t)^{\intercal}\mathbf{y}_{t}),
\end{align}
respectively, where the equality on the right-hand side of \cref{EQ:W} follows by \cref{ASS:STATIONARY}. 
Here, $f(\mathbf{x}_t)^{\intercal}\mathbf{y}_{t}$ is the welfare contribution of subject $t$ weighted according to the randomisation over treatment arms that is induced by $f$. 
We note that the optimal welfare defined by $\sup_{f \in \mathfrak{F}}W_T(f)$ agrees with the social planner's target welfare of the super-population in the static setting of treatment choice as in \citet{KT}. 

Given $(k_t)_{t=1}^{T}$---a sequence of treatment arms assigned to each subject---we define the \textit{empirical regret} and \textit{regret} as
\begin{align}
\label{EQ:R_EMP}
\hat{R}_T
& \doteq
\sup_{f\in\mathfrak{F}}\hat{W}_{T}(f)-\textstyle\sum_{t=1}^{T}y_{t}(k_t),\\
\label{EQ:R}
R_T
& \doteq
\sup_{f\in\mathfrak{F}}W_{T}(f)-\textstyle\sum_{t=1}^{T}y_{t}(k_t),
\end{align}
respectively. 
$\hat{R}_T$ is defined relative to the benchmark of the maximal empirical welfare, while $R_T$ is defined relative to the maximal average welfare. 
Since covariates, potential outcomes (and so realised outcomes), and realised treatments are all random, both $\hat{R}_T$ and $R_T$ are random. 
We assess the performance of a treatment assignment policy generating $(k_t)_{t=1}^T$ by some distributional features of empirical regret or regret. 
In particular, the performance guarantees that we provide in \Cref{SEC:GUARANTEES} for the EXP4.P algorithm are stated in terms of a uniform high-probability upper bound for the right-tail of the distribution of $\hat{R}_T$ or of $R_T$.  
The reason that we focus on a high-probability upper bound rather than the mean of the regret distribution is that bounding the mean can overlook the thick-tail phenomenon of the regret distribution for EXP-type algorithms \citep[see, for instance,][\S Chapter 12]{Lattimore_2020}.
Conversely, by noting that the mean of regret can be written as the integration of tail probabilities,  we can obtain an upper bound for the mean regret based upon the high-probability upper bounds for the tail probabilities (as long as the implementation of the algorithm does not depend upon the confidence level). 
We reiterate that the regret criterion that we consider in this paper is defined in terms of the $T$ subjects in the sample rather than in terms of the super-population of subjects, which is in contrast to the literature on best-arm identification in bandit problems.  

\section{The EXP4.P algorithm and performance guarantees} 
\label{SEC:GUARANTEES}
The EXP4.P algorithm is a learning algorithm that aims to find a best expert by efficiently balancing exploration and exploitation. 
It incorporates information about the effectiveness of previously undertaken interventions into the choice of current assignment by ascribing more weight to experts that have previously performed well. 
In this section, we introduce the EXP4.P algorithm and provide high-probability performance guarantees for the regret that it incurs. 
Our exposition depends upon whether \Cref{ASS:COMPLEXITY-FINITE} or \Cref{ASS:COMPLEXITY-VC} holds, and we separate our analysis of these two cases---modifying the EXP4.P algorithm accordingly to operate in each environment.

\subsection{The F--EXP4.P variant} \label{subsec:F-EXP4.P}
In this section, we assume that \cref{ASS:COMPLEXITY-FINITE} holds such that there there are a finite number of experts in $\mathfrak{F}$, and adapt the EXP4.P algorithm to this setting---we refer to this variant of the algorithm as F--EXP4.P.

\begin{algorithm}[F--EXP4.P; \citealp{EXP4.P}] 
\label{ALGO:F-EXP4.P} 
Input the following objects.
\begin{enumerate}[label=\roman*.]
\item A collection of tuning parameters, denoted by $(\beta,\gamma,\eta)$, such that $0 \leq \beta \leq 1$, $0 \leq \gamma \leq 1$, and $\eta \geq 0$; 
\item a finite class of experts, denoted by $\mathfrak{F}$, such that $\mathfrak{F}=\{f^{1}, \dots, f^{N}\}$ with $N = |\mathfrak{F}| < \infty$; and
\item a maximum outcome value, denoted by $M$, such that $0<M<\infty$.
\end{enumerate}
Let $q_1^i = 1/N$ for $i=1, \dots, N$. For each $t =1,...,T$, iterate
\begin{enumerate}
\item For each $k=1,...,K$, calculate a policy weight via
\begin{equation}
\label{EQ:P-REG}
p_{t}(k) 
= 
[1-\gamma]\cdot\textstyle\sum_{i=1}^{N}f_{k}^{i}(\mathbf{x}_{t})\cdot q_{t}^{i}+\gamma/K.
\end{equation} 
\item Sample $k_{t}$ from $\{1, \dots, K \}$ according to $(p_{t}(1),...,p_{t}(K))$.
\item For each $k=1,...,K$, calculate an estimate of the associated potential outcome via
\begin{equation} 
\label{EQ:IPW}
\tilde{y}_{t}(k)
= 
[\beta\cdot M^{2}+y_{t}(k_{t})\cdot 1(k_{t}=k)]/p_{t}(k).
\end{equation}
\item For each $i=1,...,N$, calculate a score via
 \begin{equation}
\tilde{s}_{t}^{i}
=
\textstyle\sum_{k=1}^{K}f_{k}^{i}(\mathbf{x}_{t})\cdot\tilde{y}_{t}(k),
\end{equation}
and a cumulative score via\footnote{%
The exponential terms in this formula can grow quickly and so, to avoid computational overflow, it is practical to normalise $\tilde{S}_{t}^{i}$ by subtracting $\textstyle\max_{\ell}\tilde{S}_{t}^{\ell}$.}
\begin{equation}
\tilde{S}_{t}^{i}
= 
\begin{cases}
\tilde{s}_{t}^{i}&\text{ if }t=1,\\
\tilde{S}_{t-1}^{i}+\tilde{s}_{t}^{i}&\text{ if }t=2,...,T.
\end{cases}
\end{equation}
\item Update the probability weights over the experts via
\begin{equation} 
\label{EQ:UPDATE-Q}
q_{t+1}^{i} 
= \exp(\eta\cdot\tilde{S}_{t}^{i})/\textstyle\sum_{\ell=1}^{N}\exp(\eta\cdot\tilde{S}_{t}^{\ell}).
\end{equation} 
\end{enumerate}
\textsc{End}\hfill\qed
\end{algorithm}

Similar to the EXP-type algorithms proposed in the machine learning literature, F--EXP4.P is premised around three key ideas.

First, F--EXP4.P administers treatment randomly according to $\mathbf{p}_t$, which we recall is a probability distribution over the treatment arms. 
This probability distribution evolves according to the outcomes that are realised in earlier periods. 
Specifically, the evolution of $\mathbf{p}_t$ is tied to the evolution of $\mathbf{q}_t$, which we recall is a probability weighting over experts; 
$\mathbf{p}_t$ assigns more weight to treatment arms that experts weighted heavily by $\mathbf{q}_t$ recommend. 
The regularisation term of $\gamma/K$ in \cref{EQ:P-REG} ensures a desirable amount of exploration, with the tuning parameter of $\gamma$ set so as to converge to zero as $T \to \infty$.  
Conversely, $\mathbf{q}_t$ is constructed based upon estimates of the average cumulative score for each expert---captured by the dependence of \cref{EQ:UPDATE-Q} upon $(\tilde{S}_{t}^{i})_{i=1}^{N}$. 
Experts with higher cumulative scores are given more weight in $\mathbf{q}_t$. 
The combined implication of $\mathbf{p}_t$ and $\mathbf{q}_t$ is that a treatment arm whose average outcome is estimated to be higher is more likely to be assigned.

Second, to construct estimates of expert performance, F--EXP4.P enacts an inverse probability weighting. 
Ignoring the regularisation term of $\beta \cdot M^2$ in \cref{EQ:IPW}, the remaining term is the observed outcome inverse weighted by the probability of assigned treatment, which provides an unbiased estimate for the average potential outcome conditional on $\mathcal{I}_t$ and $\mathbf{x}_t$. 
The regularisation term of $\beta\cdot M^2$ provides a default imputation value for the missing potential outcomes. 
\cref{THM:EXP4.P} suggests a value for $\beta$ that converges to zero---alongside the regularisation term of which it is a part---as $T \to \infty$.

Third, F--EXP4.P constructs the probability weighting over the experts so that it is proportional to the exponential tilting of estimates of their cumulative scores. 
The sensitivity of $\mathbf{q}_t$ to realisations of data is controlled by $\eta$, which we emphasise is positive;
the larger $\eta$ is, the more sensitive are the weights to experts' cumulative scores, implying that the algorithm leans more towards exploitation than exploration.
The choice of $\eta$ balances the relative strength of the exploration and exploitation motives and so affects the concentration rate of the algorithm.

We now present a welfare-regret guarantee for F--EXP4.P in terms of the empirical regret. The choices of the tuning parameters presented in the next theorem optimizes the rate of high-probability regret upper bound.  

\begin{theorem}[\citealp{EXP4.P}\S Theorem 2] \label{THM:EXP4.P}
Assume that \Cref{ASS:COMPLEXITY-FINITE} holds alongside \cref{ASS:STATIONARY,ASS:UNIFORM}.
Let $0<\delta<1$ and $\omega
\doteq
\sqrt{\ln(N/\delta)/K}$, and set 
\begin{gather}
\label{EQ:GREEKS}
\begin{aligned}
\beta
&=
\omega\cdot\sqrt{1/T}\cdot 1/M,\\
\gamma
&=
\omega\cdot\sqrt{1/T}\cdot\sqrt{\ln(N)/\ln(N/\delta)}\cdot K,
\\
\eta
&=
\omega\cdot\sqrt{1/T}\cdot\sqrt{\ln(N)/\ln(N/\delta)}\cdot 1/2M,
\end{aligned}
\end{gather}
as the parameters of the EXP4.P algorithm.
Provided that $\mathfrak{F}$ includes a randomising expert $f^{\mathrm{random}}$ that ascribes equal probability to each treatment arm (or includes a subset of experts that can be linearly combined to mimic one) and $\textstyle\max(\omega^{2},4 K\cdot\ln(N))\leq T$, then
\begin{equation}
\label{EQ:F-RLG}
\hat{R}_T
\leq
7M\cdot\omega\cdot\sqrt{K^{2}\cdot T} = 7M \sqrt{ K \cdot T \cdot \ln(N/\delta)}
\end{equation}
holds with probability at least $1-\delta$.
\end{theorem}
 
We present proof of \cref{THM:EXP4.P} in \Cref{SEC:PROOFS}.
\cref{THM:EXP4.P} extends \citet[][\S Theorem 2]{EXP4.P}, which requires that the potential outcomes be contained in the unit interval---rather than bounded and non-negative, as we assume. 
This extension modifies the choice of tuning parameters from that of \citet[][\S Theorem 2]{EXP4.P}.  
The maximal regret bound that we obtain then differs from that of \citet[][\S Theorem 2]{EXP4.P} in that the regret that we obtain is scaled by $M$, and we include this as an additional parameter.
We emphasise that our maximal regret bound---and the algorithm itself---coincides with that of \citet[][\S Theorem 2]{EXP4.P} if the potential outcomes are contained in the unit interval.

\begin{remark} \label{REMARK:INVARIANCE}
We note that the assignment policies that are delivered by F--EXP4.P (with tuning parameters chosen according to \cref{EQ:GREEKS}) are not invariant to the addition of a constant to the outcome variable. 
This is because the inverse probability weighted estimate of the expert's score (denoted by $\tilde{s}_t^{i}$) is not equivariant to additive constants. 
Accordingly, in those situations in which outcomes are observed to take negative values, the allocations produced by \cref{ALGO:F-EXP4.P} can be sensitive to the choice of constant that is added to outcomes so as to satisfy the nonnegativity requirement of \cref{ASS:UNIFORM}.  
\end{remark}

\subsection{The VC--EXP4.P variant} \label{subsec:VC-EXP4}

The standard setting in policy learning is one in which $\mathfrak{F}$ contains infinitely many treatment assignment rules.
F--EXP4.P restricts the class of treatment assignment rules to be finite (i.e., only a finite number of individualised assignment rules are allowed), and so is incompatible with the standard setting.
In this section, we assume that \cref{ASS:COMPLEXITY-VC} holds such that there are infinitely many deterministic (i.e., non-random) experts in $\mathfrak{F}$ with complexity controlled by a finite VC dimension, and adapt the EXP4.P algorithm to this setting---we refer to this variant of the algorithm as VC--EXP4.P.

\begin{algorithm}[VC--EXP4.P; \citealp{EXP4.P}] \label{ALGO:VC-EXP4.P}
Input the following objects.
\begin{enumerate}[label=\roman*.]
\item A duration for the coarsening phase, denoted by $\tau$, such that $\tau\in\mathbb{R}_{++}$ and $\tau<T$;
\item a class of experts, denoted by $\mathfrak{F}$, of finite VC dimension; and
\item a maximum outcome value, denoted by $M$, such that $0<M<\infty$.
\end{enumerate}
Generate treatment assignments by implementing the following algorithm:
\begin{enumerate}
\item For $t=1,...,\lceil\tau\rceil$, sample $k_{t}$ randomly from the uniform distribution on $\{1,...,K\}$.

\item Coarsen $\mathfrak{F}$ to $\mathfrak{G} \subset \mathfrak{F}$ such that for any $f \in \mathfrak{F}$, there exists exactly one $g \in \mathfrak{G}$ satisfying $f(\mathbf{x}_{t}) = g(\mathbf{x}_{t})$ for all $t=1,...,\lceil\tau\rceil$.
\item Let $N=|\mathfrak{G}| - 1$ and implement F--EXP4.P over $t=\lceil\tau\rceil+1,...,T$ using $\mathfrak{G} \cup f^{\mathrm{random}}$ as the set of experts.
\end{enumerate}

\textsc{End}\hfill\qed
\end{algorithm}


The key innovation of VC--EXP4.P is the initial refinement of $\mathfrak{F}$ to $\mathfrak{G}$, which occurs during an initial coarsening phase.
Once $\mathfrak{G}$ is formed, the algorithm proceeds to a run phase during which F--EXP4.P is enacted on $\mathfrak{G} \cup f^{\mathrm{random}}$. The reason that we append $f^{\mathrm{random}}$ to the coarsened class is to meet the assumption for the regret guarantee of F-EXP4.P shown in \cref{THM:EXP4.P}. 
Whether this coarsening is straightforward to implement or not depends upon how $\mathfrak{F}$ is specified. 
For instance, when $\mathfrak{F}$ comprises assignment rules based upon linear eligibility scores, we present a feasible algorithm for hyperplane arrangements to refine $\mathfrak{F}$ to $\mathfrak{G}$. 

We now present a welfare-regret guarantee for VC--EXP4.P that holds for particular choices of the tuning parameters and duration of the coarsening phase.  

\begin{theorem}[\citealp{EXP4.P}\S Theorem 5] \label{THM:EXP4.P-VC}
Assume that \cref{ASS:COMPLEXITY-VC} holds alongside \cref{ASS:STATIONARY,ASS:UNIFORM}. 
Let $0<\delta<1$ and set 
\begin{gather}
\label{EQ:GREEKS-REVISED}
\begin{aligned}
\beta
&=
\omega\cdot\sqrt{1/[T-\lceil\tau\rceil]}\cdot 1/M,\\
\gamma
&=
\omega\cdot\sqrt{1/[T-\lceil\tau\rceil]}\cdot\sqrt{\ln(N)/\ln(N/\delta)}\cdot K,
\\
\eta
&=
\omega\cdot\sqrt{1/[T-\lceil\tau\rceil]}\cdot\sqrt{\ln(N)/\ln(N/\delta)}\cdot 1/2M,
\end{aligned}
\end{gather}
as the parameters of the EXP4.P algorithm given 
\begin{equation}
\label{EQ:TAU}
\tau
=
\sqrt{T\cdot [2D\cdot\ln(T\cdot e/D)+\ln(3/\delta)]},
\end{equation}
as the duration of the coarsening phase.
Provided that $D\ll T$,\footnote{%
For the EXP4.P algorithm to be well-defined, $\mathfrak{F}$ or $\mathfrak{G}$ cannot be too rich.
The condition that $\textstyle\max(\omega^{2},4 K\cdot\ln(N))\leq T$ is necessary to ensure this when $\mathfrak{F}$ is finite.
When $\mathfrak{F}$ is infinite, however, we require not only that this condition holds for the $\mathfrak{G}$ that is obtained from it but that $\tau<T$.
Since $\omega^{2}\leq\tau$, it is necessary that 
\begin{equation}
\label{EQ:MINIMUM-T-INFINITE}
\textstyle\max(8\ln(N),2\ln(N)+\ln(2/\delta))
\leq 
T
\end{equation}
holds, which, under $N=|\mathfrak{G}|$, translates to 
\begin{equation}
\label{EQ:MINIMUM-T-EXPLICIT}
T
\in
\{t : \textstyle\max(8D\cdot\ln(\tau\cdot e/D),2D\cdot\ln(\tau\cdot e/D)+\ln(2/\delta))\leq t\,|\,\tau=\sqrt{t\cdot [2D\cdot\ln(t\cdot e/D)+\ln(2/\delta)]}\}
\end{equation}
by Sauer's Lemma.
For \cref{EQ:MINIMUM-T-INFINITE} to hold generally then, the condition that $D\lll T$ is sufficient (i.e., the VC dimension is small relative to the time horizon).} 
then, with a universal constant $c>0$, 
\begin{equation}
\label{EQ:VCREGRET}
R_T
\leq 
cM\cdot\sqrt{\lceil\tau\rceil^{2}+T\cdot\ln(3/\delta)}
\end{equation}
holds with probability at least $1-\delta$.
\end{theorem}

We present proof of \cref{THM:EXP4.P-VC} in \Cref{SEC:PROOFS}. 
An important implication of controlling the complexity of (deterministic) experts in $\mathfrak{F}$ by limiting the VC dimension is that, given $(\mathbf{x}_{t})_{t=1}^{\lceil\tau\rceil}$, we can construct $\mathfrak{G} = \{g: \mathcal{X} \to \{0, 1 \} \}$ to represent the remaining experts in $\mathfrak{F}$ in the sense that there exists exactly one $g \in \mathfrak{G}$ satisfying $f(\mathbf{x}_{t}) = g(\mathbf{x}_{t})$ for all $t=1,...,\lceil\tau\rceil$.
The cardinality of $\mathfrak{G}$ can be much smaller than $2^{\tau}$, the cardinality of binary functions mapping $\tau$ points. 
\Cref{THM:EXP4.P-VC} implies that choosing $\tau$ to be of the order of $ O(\sqrt{T\cdot D\cdot \ln(T)})$ gives a sufficient degree of coarsening for the run phase to have a high-probability regret bound of the order of $O(\sqrt{T\cdot D\cdot \ln(T)})$.  

An important and practically relevant class of experts is the Linear Eligibility Score (LES) class \citep{KT}:
\begin{equation}
\mathfrak{L}_J \doteq \{ ( 1 ( (1,\mathbf{x}^{\intercal} ) \beta < 0 ), 1 ( (1,\mathbf{x}^{\intercal} ) \beta \geq 0 ))^{\intercal}  : \beta \in \mathbb{R}^{J+1} \},
\end{equation}
which corresponds to the class of 0-1 assignments spanned by hyperplane partitions in $\mathcal{X} \subset \mathbf{R}^J$.
If $\mathfrak{F}$ is an LES class, we can improve upon \cref{EQ:VCREGRET} by providing a tighter bound on the cardinality of $\mathfrak{G}$.
Define, for $t \geq 2$
\begin{equation}
\label{EQ:HARDINGNUMBER}
\mathrm{Harding}(t,J)
\doteq
2\textstyle\sum_{j=0}^{J}\textstyle\binom{t-1}{j},
\end{equation}
as per \citet{EF}.
We note that $\mathrm{Harding}(t,J)$ is the maximum number of distinct signed partitions of $t$ points in $\mathbb{R}^{J+1}$---the ambient space---that can be induced by hyperplanes.\footnote{%
\citet{EF} in fact determines the maximum number of distinct partitions of $t$ points in $\mathbb{R}^{J+1}$ that can be induced by hyperplanes;
the maximum number of distinct signed partitions is obtained trivially by doubling the maximum number of distinct partitions.} 
As such, \cref{EQ:HARDINGNUMBER} corresponds to the maximum number of linear assignment rules that induce distinct allocations of $t$ observations over the two treatment arms.
It is straightforward to show that \cref{EQ:HARDINGNUMBER} is less than the corresponding bound on the cardinality of $\mathfrak{G}$ that is implied by Sauer's Lemma \citep{SAUER,SHELAH,VC}---which is central to our derivation of \cref{THM:EXP4.P-VC}---when $\mathfrak{F}=\mathfrak{L}_{J}$.

To implement improvement of \cref{EQ:VCREGRET} then, we leverage \cref{EQ:HARDINGNUMBER} to derive a tighter bound on the maximal cardinality of $\mathfrak{F}$ when this coincides with the LES class.

\begin{theorem} \label{THM:EXP4.P-LES}
Assume that \cref{ASS:COMPLEXITY-VC} holds alongside \cref{ASS:STATIONARY,ASS:UNIFORM} and, moreover, that $\mathfrak{F}=\mathfrak{L}_{J}$. 
Let $0<\delta<1$ and set 
\begin{gather}
\label{EQ:GREEKS-REVISED FOR LES}
\begin{aligned}
\beta
&=
\omega\cdot\sqrt{1/[T-\tau]}\cdot 1/M,\\
\gamma
&=
\omega\cdot\sqrt{1/[T-\tau]}\cdot\sqrt{\ln(N)/\ln(N/\delta)}\cdot K,
\\
\eta
&=
\omega\cdot\sqrt{1/[T-\tau]}\cdot\sqrt{\ln(N)/\ln(N/\delta)}\cdot 1/2M,
\end{aligned}
\end{gather}
as the parameters of the EXP4.P algorithm given 
\begin{equation}
\label{EQ:TAULES}
\tau
=
\sqrt{T\cdot[2\ln\circ\mathrm{Harding}(T,J)+\ln(3/\delta)]},
\end{equation}
as the duration of the coarsening phase.
Provided that $J+1\ll T$,
\begin{equation}
\label{EQ:LESREGRET}
R_{T}
\leq 
cM\cdot\sqrt{\left\lceil\tau\right\rceil^{2}+T\cdot\ln(3/\delta)}
\end{equation}
holds with probability at least $1-\delta$ with a univesal constant $c>0$.
\end{theorem}

Coarsening the LES class can be formulated as a cell enumeration problem and achieved using an incremental enumeration algorithm---outlined below, alongside some practical recommendations.

\subsection{Coarsening the class of linear eligibility assignment rules} \label{SEC:HYPERPLANES}
Suppose that $\mathfrak{F}=\mathfrak{L}_{J}$ so that $f \in \mathfrak{F}$ can be indexed by $\beta\in\mathbb{R}^{J+1}$.
The fundamental insight that we build upon to coarsen $\mathfrak{F}$ to $\mathfrak{G}$ is that, for $f^{i}\in\mathfrak{G}$ and $f^{\ell}\in\mathfrak{G}$, there must exist some $\mathbf{x}_{t}$ in $(\mathbf{x}_{t})_{t=1}^{\lceil\tau\rceil}$ such that
\begin{equation}
\label{EQ:COARSE-FUNDAMENTAL}
\mathrm{Sign}((1,\mathbf{x}_{t}^{\intercal})\beta^{i})
\neq
\mathrm{Sign}((1,\mathbf{x}_{t}^{\intercal})\beta^{\ell}).
\end{equation}
In words, that we can find a subject (during the coarsening phase) for whom $f^{i}$ and $f^{\ell}$ recommend different treatments.

Mathematically, \cref{EQ:COARSE-FUNDAMENTAL} is equivalent to $\beta_{i}$ and $\beta_{\ell}$ being located on opposite sides of the hyperplane defined by 
\begin{equation}
\label{EQ:HYPERPLANE}
H_{t}
\doteq
\{\beta : (1,\mathbf{x}_{t}^{\intercal})\beta=0\}.
\end{equation}
The problem of constructing the coarse class for the LES class can be seen as a cell enumeration problem.
That is, the problem of determining which cells---identified by a label (i.e., the concatenation of a sign for each subject in the sample according to whether the left- or right-hand side of \cref{EQ:COARSE-FUNDAMENTAL} is true) and equivalent to a partition of the parameter space of $\beta$---are compatible with a given hyperplane arrangement, and calculating a point in the interior of each cell.

The cell enumeration formulation highlights that $\mathfrak{G}$ is not unique.
If $\beta^{i}$ and $\beta^{\ell}$ both induce the same label in the sample then $\mathfrak{G}$ can include $\beta^{i}$ or $\beta^{\ell}$ but not both, and are said to belong to a common equivalence class---or cell.
The invariance of treatment allocations in the same equivalence class is also exploited by \citet[][see also \citealp{rosen2019finite}]{JP} in the closely-related problem of maximum score estimation.

The incremental enumeration algorithm introduced by \citet[building upon earlier work by \citealp{avis1996reverse,sleumer1998output}]{RS} and improved upon by \cite{GU} is an efficient way to construct $\mathfrak{G}$.
It considers the enumeration problem one hyperplane at a time, exploiting the similarity of neighbouring cells.
The number of operations that the algorithm involves is proportional to the number of cells that are compatible with a hyperplane arrangement, which \citet{BUCK} establishes is less than $\tau^{J}$, and far fewer than the $2^{\tau}$ partitions that a na\"{i}ve \textit{brute force} algorithm operates over and that is infeasible for even moderately-sized samples.

Before discussing how the incremental enumeration algorithm works, however, we clarify what we mean by \emph{cell} and by \emph{label}.
To facilitate this clarification, imagine that we (are able to and do) plot a hyperplane of the form that is given in \cref{EQ:HYPERPLANE} for each subject in the sample.
A cell is the intersection of half-spaces defined by a given hyperplane arrangement.
A label is a consistent identifier of whether a point is located in the positive or negative half-spaces defined by a hyperplane arrangement.
For instance, by recording whether a cell is above or below each hyperplane in the sense of \cref{EQ:COARSE-FUNDAMENTAL}, we can identify it by a sequence of pluses and minuses, which we refer to as a label.
A na\"{i}ve consideration of the problem then is to consider a label---of which there are $2^{\tau}$ possible permutations of pluses and minuses---and to determine whether such a label identifies a non-empty cell.
Given that each cell is defined as the intersection of half-spaces, this amounts to solving a linear programme in which the objective is to maximise a slackness variable $0 \leq r \leq 1$ subject to the constraints of the form;
\begin{equation}
\label{EQ:LP}
\mathrm{Constraint}_{t}
=
\begin{cases}
(1,\mathbf{x}_{t}^{\intercal})\beta_{-}-(1,\mathbf{x}_{t}^{\intercal})\beta_{+}+r\cdot\|(1,\mathbf{x}_{t}^{\intercal})\|_{2}\leq 0,
&
\text{ if } \mathrm{Sign}((1,\mathbf{x}_{t}^{\intercal})\beta)=+,
\\
(1,\mathbf{x}_{t}^{\intercal})\beta_{+}-(1,\mathbf{x}_{t}^{\intercal})\beta_{-}+r\cdot\|(1,\mathbf{x}_{t}^{\intercal})\|_{2}\leq 0,
&
\text{ if } \mathrm{Sign}((1,\mathbf{x}_{t}^{\intercal})\beta)=-,
\end{cases}
\end{equation}
where $\beta_{+}\in\mathbb{R}_{+}^{J+1}$ and $\beta_{-}\in\mathbb{R}_{+}^{J+1}$ are the positive and negative parts of $\beta$, respectively, in the sense that $\beta=\beta_{+}-\beta_{-}$ holds, and $(\beta_{+},\beta_{-})$ are choice variables in the linear programme. 

The incremental enumeration algorithm exploits the fact that if there does not exist a feasible solution to the linear programme for $t$ subjects (i.e., the linear programme with constraints of the form that is given in \cref{EQ:LP}---one constraint for each $\ell=1,...,t$) for a given label, then there is no feasible solution to the linear programme for $t+1$ subjects that appends this label by a plus or a minus.
Put simply, empty cells cannot be divided.

We now present a (deliberately) simple description of the steps of the incremental enumeration algorithm (see \citealp{RS} for a more detailed description).

\begin{algorithm}[Incremental enumeration; \citealp{RS}] \label{ALGO:IE}
The user is required to input the following objects.
\begin{enumerate}[label=\roman*.]
\item A duration for the coarsening phase, denoted by $\tau$, such that $\tau\in\mathbb{R}_{++}$ and $\tau<T$; and
\item a LES class of experts $\mathfrak{L}_J$ of finite VC dimension $D = J+1$.
\end{enumerate}
For $t=1,...,\lceil\tau\rceil$, iterate.
\begin{enumerate}[label=\alph*.]
\item If $t=1$ then propose $+$ and $-$ as labels;
else append $+$ and $-$ separately to any existing labels comprising $t-1$ pluses and minuses.
\item Construct and solve a linear programme with constraints of the form that is given in \cref{EQ:LP}---one constraint for each $\ell=1,...,t$---for each proposed label.
\item If the linear programme associated with a proposed label has a feasible solution then keep this label and store the solution; 
else disregard the proposed label.
\end{enumerate}
For each stored label, construct and solve a linear programme with constraints of the form that is given in \cref{EQ:LP}---one constraint for each $\ell=1,...,t$---and return $\mathfrak{G}$ as the collection of $\beta=\beta_{+}-\beta_{-}$ that are found as part of the solution.

\noindent\textsc{End}\hfill\qed
\end{algorithm}

Since the total number of labels that can be proposed at any step of \Cref{ALGO:IE} is twice the number of labels that are carried forward from the previous step, the total number of labels that need to be checked is given by \cref{EQ:HARDINGNUMBER}.
Provided that a label is admitted, then a natural witness to the associated cell is its Chebyshev centre;\footnote{%
The Chebyshev centre of a bounded convex set is defined as the origin of the largest possible circle that can be drawn that is entirely enclosed within that set;
by specifying an upper limit on the radius of the circle, we can compute something akin to the Chebyshev centre for a possibly unbounded convex set.
We emphasise that cells need not be unbounded, with notable examples being $+$ and $-$ (i.e., the cells that are compatible with a single hyperplane).}
the solution to the linear programme for $t$ subjects is the pseudo-Chebyshev centre of a cell and is, incidentally, also the witness to its division (i.e., any non-empty cell that is obtained from it upon the addition of another constraint).

\begin{figure}[t]
\centering
\includegraphics{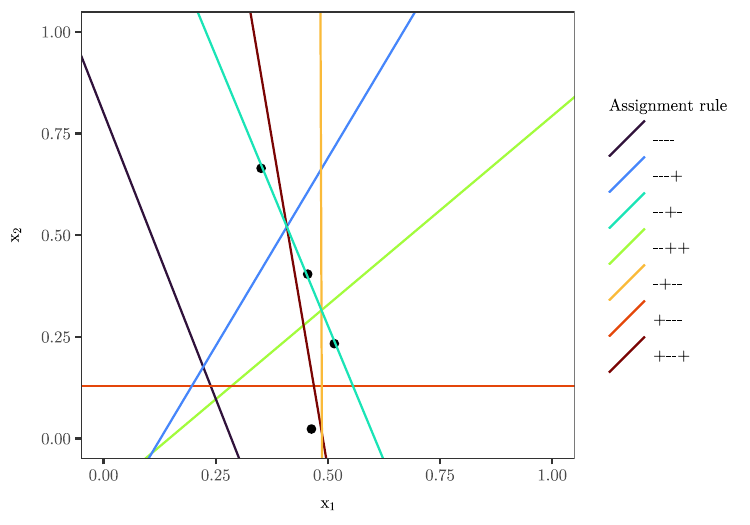}
\caption{A sufficient characterisation of admissible treatment rules for $\tau = 4$ observations.}
\label{FIG:GENERAL}
\end{figure}

\cref{FIG:GENERAL} illustrates the output of the incremental enumeration algorithm for four randomly generated points.
Every admissible cell is the mirror image of another;
it is, therefore, sufficient to characterise the full set of admissible cells by excluding all cells that are mirror images of another cell in the characterisation.
We exploit this property in \cref{FIG:GENERAL} to limit the number of treatment rules that we need to state.
The generated points are compatible with a total of 14 cells, which is fewer than the 16 cells that na\"{i}ve calculation suggests.

\section{Linear experiments}
\label{SEC:EXERCISES}
To provide some insight into the practical implementation and performance of VC--EXP4.P,\footnote{%
We use three Intel Core i7-8700 CPU @ 3.20GHz computers equipped with 32GB RAM running Ubuntu 20.04.6 LTS (Focal Fossa) or Windows 10 Enterprise LTSC and R 4.3.0.} 
we create several artificial datasets, run VC--EXP4.P with the LES class of assignments, and assess its welfare-performance. 
\subsection{Design}
We construct 1,000 artificial datasets, each comprising 1,000 realisations of a collection of two covariates---which we denote by $\mathbf{x}_{t}=(x_{t,1},x_{t,0})^{\intercal}\in\mathbb{R}^{2}$, as before---and of a collection of two private shocks---which we denote by $\mathbf{u}_{t}=(u_{t,1},u_{t,0})^{\intercal}\in\mathbb{R}^{2}$.
The covariates and private shocks---which play the role of underlying heterogeneity---together generate a subject's potential outcomes. 
We fix the realisations of the covariates across the datasets (i.e., $\mathbf{x}_{t}$ is the same in every dataset) but allow the realisations of the private shocks to vary;
by also fixing $\tau$, we can use the same $\mathfrak{G}$ throughout our simulations to significantly reduce computation time.
We relate the covariates and unobserved heterogeneity to the potential outcomes via a log-normal specification. 
We maintain log-normality throughout the remainder
of \Cref{SEC:EXERCISES} only, assuming that the covariates are generated from a uniform distribution on the unit square.
Motivating our choice of specification is the appropriateness of a log-normal distribution as an approximation of income or other wealth-related indices---common measures of welfare in treatment choice applications---and the non-negativity of log-normal random distributions---in accordance with the non-negativity requirement of \cref{ASS:UNIFORM}.
The log-normality assumption does not, however, satisfy the boundedness requirement of \cref{ASS:UNIFORM}. 
To implement VC--EXP4.P, we use the first $\tau$ observations to both compute $\mathfrak{G}$ and to inform our choice of the tuning parameters of the F--EXP4.P-step that occurs during the run phase.

\begin{assumption}[Log-normality]
\label{ASS:LOGNORMALITY}
The relationship between the covariates and unobserved heterogeneity, and the potential outcomes is governed by
\begin{equation}
\label{EQ:DIFFICULTY}
\begin{aligned}
y_{t,1}
&=
\exp(x_{t,1}-x_{t,2}+u_{t,1}-\sigma^{2}/2),\\
y_{t,0}
&=
\exp(u_{t,0}-\sigma^{2}/2),
\end{aligned}
\end{equation}
where $\mathbf{u}_{t}\sim\mathrm{Normal}((0,0)^{\intercal},\sigma^{2}\cdot\mathbb{I}_{2})$.
\end{assumption}

We note that, aside from its interpretation as a standard deviation, $\sigma$ controls the signal-to-noise ratio under \cref{ASS:LOGNORMALITY}, and is, therefore, related to what we regard as the degree of difficulty of learning the optimal policy of an artificial dataset.

The first-best optimal assignment rule under \cref{ASS:LOGNORMALITY} is given by $1\left\{ x_{t,1}\geq x_{t,2}\right\}$. 
This rule belongs to the LES class.

A property of the EXP4.P algorithm---and other exponential tilting procedures---is that it more heavily updates the current policy when the realised outcome is large.
Compressing the distribution of effects (i.e., the differences between the potential outcomes) is, in theory, likely to decrease the convergence rate of the EXP4.P algorithm to whatever assignment rule is best-in-class.

Our intention in normalising the potential outcomes by $\sigma^2/2$ is to ensure comparability of the means of the potential outcomes for different standard deviations of the unobserved heterogeneity. 
To guarantee that \cref{ASS:UNIFORM} holds, we normalise the potential outcomes by dividing \cref{EQ:DIFFICULTY} by the maximum value of the potential outcomes that is realised for a given standard deviation of the unobserved heterogeneity,\footnote{%
For these experiments, we leverage our position as the oracle to set $M$ equal to the maximum value of both potential outcomes across all of the artificial datasets. 
In practice, this is not feasible and so we suggest using the first $\tau$ observations to estimate some maximal value, or else to rely on insight or economic theory.}
but reverse this normalisation prior to presenting the results of our experiments to maintain comparability.

To vary the difficulty of learning, we manipulate the variance of the unobserved heterogeneity in \cref{EQ:DIFFICULTY}.
We then measure the corresponding difficulty via the population probability of misclassification under the optimal rule.
\begin{figure}[t]
\centering
\includegraphics{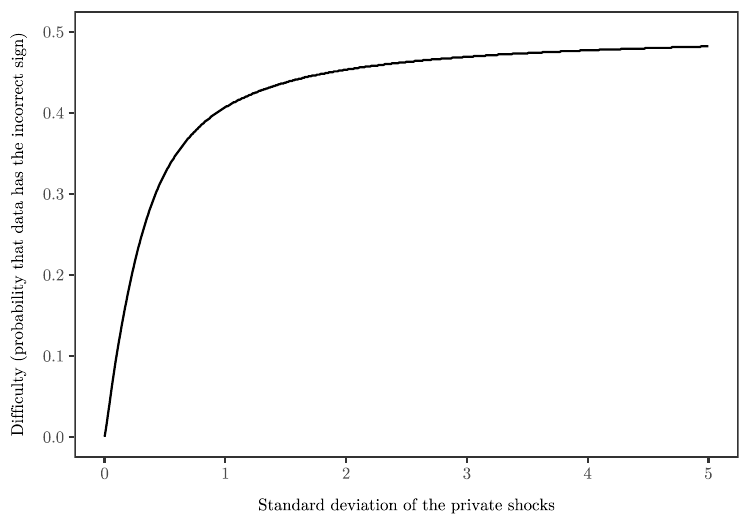}
\caption{The difficulty of \cref{EQ:DIFFICULTY} computed using Monte Carlo approximation.}
\label{FIG:DIFFICULTY}
\end{figure}
\begin{equation}
\label{EQ:MISCLASSIFICATION}
\text{Difficulty}
=
\Pr(\textrm{Sign}(y_{t,1}-y_{t,0})\neq\textrm{Sign}(x_{t,1}-x_{t,2})).\\
\end{equation}
\cref{FIG:DIFFICULTY} plots the relationship between the standard deviation of the unobserved heterogeneity and the measure of difficulty;
the black line is the theoretical difficulty implied by \cref{EQ:MISCLASSIFICATION}; 
and the orange band is the range of actual difficulty across the various artificial datasets (i.e., replacing the probability in \cref{EQ:MISCLASSIFICATION} with the average of an indicator of the specified event).

\subsection{Implementation}
We restrict attention to LES rules with two covariates and set $\mathfrak{F} = \mathfrak{L}_2$.
To coarsen $\mathfrak{L}_2$, we utilise the incremental enumeration algorithm described in \Cref{SEC:HYPERPLANES}.
We find that 31,154 assignment rules are compatible with the first 177 realisations of the covariates (the duration of the coarsening phase suggested by \cref{EQ:TAULES} given \cref{EQ:GREEKS}),\footnote{%
As a guide, our implementation of the incremental enumeration algorithm completes in seven minutes on our computer system.} which we emphasise are constant across all 1,000 artificial datasets---we only vary the unobserved heterogeneity and so the potential outcomes.
For comparison, this is the same number of assignment rules that \citet{EF} suggests could be admitted. 

To assess whether the coarsening phase generate a sizeable welfare loss, it is useful to know whether the coarse class contains the optimal rule, or assignment rules that are close to it.
We find that the coarse class does not contain the optimal rule, but does contain several assignment rules that are close to it.
In particular, the coarse class contains an assignment rule that allocates only one individual (out of 1,000) differently to the oracle rule.

To draw meaningful conclusions about the performance of the EXP4.P algorithm, we also implement two other simple estimators whose performance we can use as a benchmark for comparison.
We implement (i.) the optimal rule through $t= 1, \dots, 1000$---what we refer to as the oracle rule; 
and (ii.) randomisation over the first 177 observations followed by the best-performing deterministic assignment rule in $\mathfrak{G}$ according to the empirical welfare criterion of \citet{KT}---what we refer to as $\tau$-EWM.

\subsection{Results}

\begin{figure}
\centering
\includegraphics{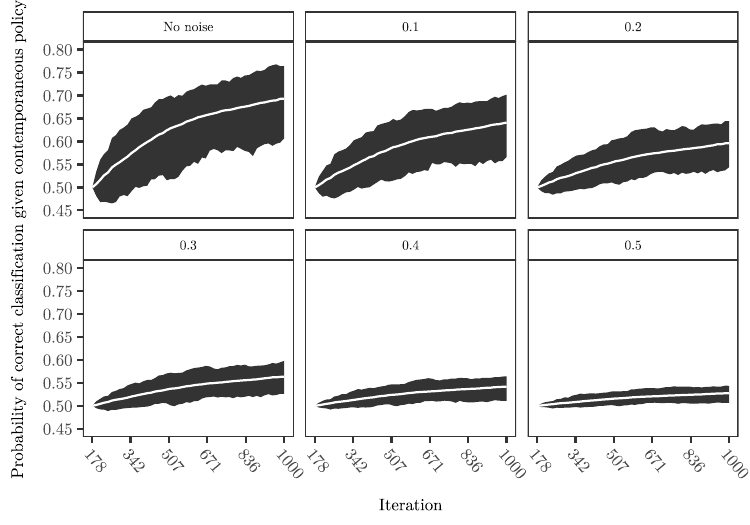}
\caption{The performance of VC--EXP4.P as the standard deviation of the unobserved heterogeneity increases.}
\label{FIG:RATIO}
\end{figure}
\begin{figure}
\centering
\includegraphics{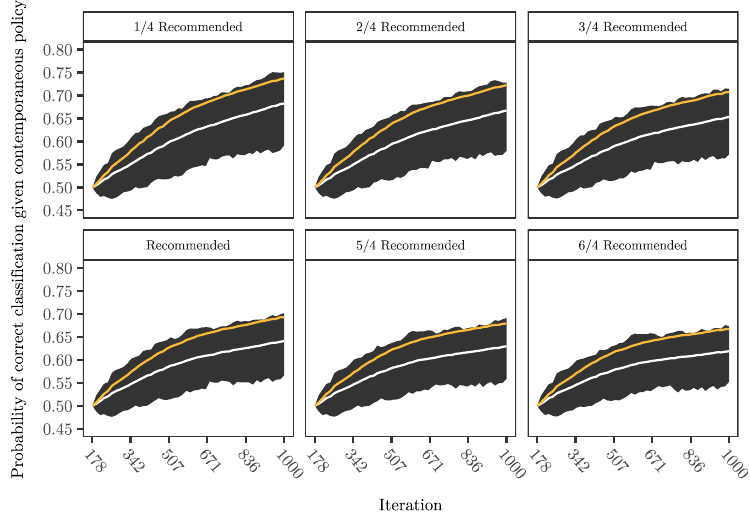}
\caption{The performance of VC--EXP4.P as $\beta$ varies given $\sigma=0.1$.}
\label{FIG:BETA}
\end{figure}
\begin{figure}
\centering
\includegraphics{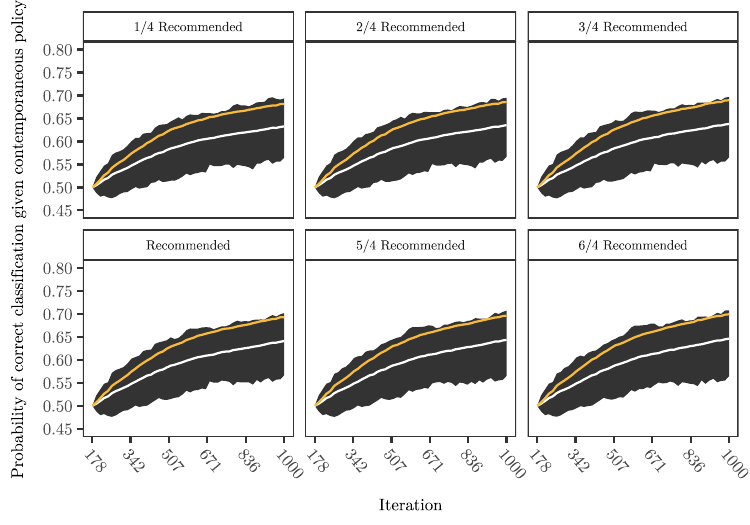}
\caption{The performance of VC--EXP4.P as $\gamma$ varies given $\sigma=0.1$.}
\label{FIG:GAMMA}
\end{figure}
\begin{figure}
\centering
\includegraphics{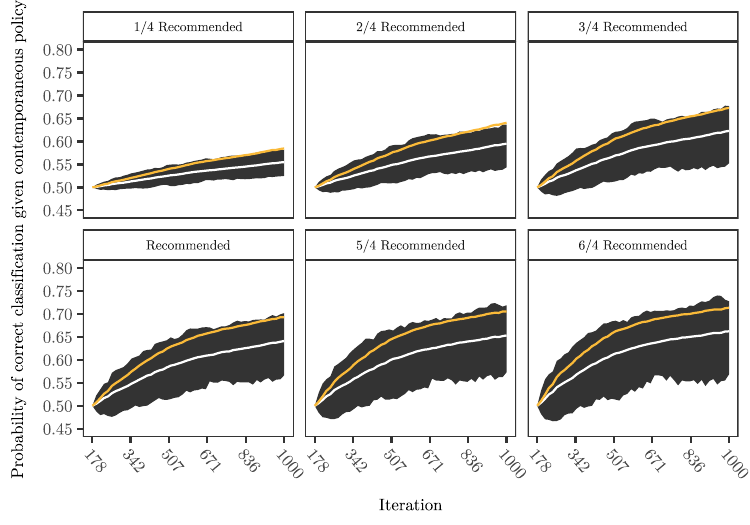}
\caption{The performance of VC--EXP4.P as $\eta$ varies given $\sigma=0.1$.}
\label{FIG:ETA}
\end{figure}

We conduct numerous experiments that vary the standard deviation of the unobserved heterogeneity or the tuning parameters of the EXP4.P algorithm.
We set $\delta$ to the standard $95\%$-level

We capture the behaviour of the EXP4.P algorithm by computing the probability that the social planner's chosen policy enacts the same intervention as the optimal assignment rule. 
This probability can be computed for every observation and is induced by $\mathbf{p}_t$---the probability distribution over the treatment arms---underlying the assignment probability of the EXP4.P algorithm.  
A higher chance of correct classification (i.e., coincidence of the treatment assignment of VC--EXP4.P and that of the optimal rule) leads to a lower regret;
how much and how quickly the probability of correct classification increases over time is informative as to the performance of the EXP4.P algorithm. 
We plot the probability of correct classification as \cref{FIG:RATIO,FIG:BETA,FIG:GAMMA,FIG:ETA}.

In \cref{FIG:RATIO,FIG:BETA,FIG:GAMMA,FIG:ETA}, the black ribbon captures the sample probability that the policy enacts the same intervention as the optimal rule in each of the 1,000 artificial datasets that we construct;
the white line that bisects the ribbon is the median probability amongst these datasets.
In \cref{FIG:BETA,FIG:GAMMA,FIG:ETA}, the orange line that dissects the ribbon is the corresponding median probability when the standard deviation of the unobserved heterogeneity is set to zero and is included for comparison.

We observe that the planner's chosen policy is more likely to coincide with the treatment assignment of the optimal rule over time but that the magnitude and rate of this improvement largely depends upon the difficulty and the choice of tuning parameters.

First, we investigate how changing the standard deviation of the unobserved heterogeneity affects performance.
In each panel of \cref{FIG:RATIO} (from left to right) we increase the standard deviation of the unobserved heterogeneity, holding fixed the tuning parameters of the EXP4.P algorithm at the level recommended by \cref{THM:EXP4.P}.
We emphasise that although each dataset reveals identical sequences of the covariates, at every iteration both the intervention (even for an identical policy) and the potential outcomes that are realised differ and it is this attribute that generates the distribution of probabilities that is captured by the black ribbon.
As is to be expected, as the standard deviation of the unobserved heterogeneity increases,
the EXP4.P algorithm finds it increasingly difficult to learn an optimal policy.
The difficulty corresponding to each panel is zero, 10.1\%, 18.2\%, 24.5\%, 29.1\%, and 32.5\%, respectively. The EXP4.P algorithm struggles to show even minimal signs of convergence for comparatively high standard deviations of the unobserved heterogeneity.

Second, we investigate how changing $\beta$ affects the performance of the EXP4.P algorithm.
In each panel of \cref{FIG:BETA} (from left to right) we scale $\beta$ relative to its recommended value in \cref{THM:EXP4.P}, holding fixed the other tuning parameters of the EXP4.P algorithm at the level recommended by \cref{THM:EXP4.P}.
The exploration motive is increasing in $\beta$, which governs how much to score unrealised treatment arms (and so too the experts that recommend these treatment arms).
\cref{FIG:BETA} suggests that the EXP4.P algorithm converges more slowly to the optimal rule as $\beta$ increases.

Third, we investigate how changing $\gamma$ affects the performance of the EXP4.P algorithm.
In each panel of \cref{FIG:GAMMA} (from left to right) we scale $\gamma$ relative to its recommended value in \cref{THM:EXP4.P}, holding fixed the other tuning parameters of the EXP4.P algorithm at the level recommended by \cref{THM:EXP4.P}. 
The exploration motive is increasing in $\gamma$, which governs the willingness to randomise versus adopting expert recommendations (weighted by the expert weights).
\cref{FIG:GAMMA} suggests that the EXP4.P algorithm is not particularly sensitive to the choice of $\gamma$. 

Fourth, we investigate how changing $\eta$ affects the performance of the EXP4.P algorithm.
In each panel of \cref{FIG:ETA} (from left to right) we scale $\eta$ relative to its recommended value in \cref{THM:EXP4.P}, holding fixed the other tuning parameters at the level recommended by \cref{THM:EXP4.P}.
The exploitation motive is increasing in $\eta$, which governs the sensitivity of the assignment probability to realised outcomes.
\cref{FIG:ETA} suggests that the EXP4.P algorithm is particularly sensitive to the choice of $\gamma$---more so than to the values of the other tuning parameters.
The median probability of correct classification is increasing in $\eta$ (i.e., the white line reaches a higher maximum), although this is accompanied by an increase in the variance of the correct classification sample probability (i.e., the black ribbon widens).
This trade-off between the average and variance of performance resembles the bias-variance trade-off that commonly appears in nonparametric estimation for the choice of a smoothing parameter. 

\begin{table}
\centering
\caption{The average welfare-level attained by various estimators as the standard deviation of the unobserved heterogeneity increases.}
\label{TAB:RATIO}
\begin{tabularx}{\textwidth}{l*{8}C}
\hline
{Estimator}							& \multicolumn{6}{c}{$\sigma$}\\
								\cmidrule{2-7}
								& No noise	& 0.1		& 0.2		& 0.3		& 0.4		& 0.5\\
\hline\\
Oracle rule							& 2.001 	& 2.001	& 2.000	& 2.000	& 2.000 	& 1.999\\
\\
$\tau$-EWM			& 1.891 	& 1.891 	& 1.891	& 1.890	& 1.886	& 1.884\\
\\
VC--EXP4.P 							& 1.782 	& 1.761 	& 1.745 	& 1.733 	& 1.726	& 1.721\\
\quad Coarsening phase					& 1.707	& 1.707	& 1.708	& 1.708	& 1.709	& 1.709\\
\quad Run phase					& 1.798	& 1.772	& 1.753	& 1.739	& 1.729	& 1.723\\
\hline
\end{tabularx}
\end{table}
\begin{table}
\centering
\caption{The average welfare-level attained by various estimators as the parameters of the EXP4.P algorithm vary given $\sigma=0.1$.}
\label{TAB:BETA}
\begin{tabularx}{\textwidth}{l*{8}C}
\hline
{Estimator}							& \multicolumn{6}{c}{Tuning parameter value}\\
								\cmidrule{2-7}
								& 1/4 Rec. 	& 2/4 Rec. 	& 3/4 Rec. & Rec. & 5/4 Rec. & 6/4 Rec.\\
\hline\\
Oracle rule							& 2.001 	& 2.001 	& 2.001	& 2.001	& 2.001 	& 2.001\\
\\
$\tau$-EWM			& 1.890 	& 1.890	& 1.890	& 1.890	& 1.890	& 1.890\\
\\
VC--EXP4.P ($\beta$)					& 1.771	& 1.767	& 1.764	& 1.761	& 1.758	& 1.755\\
\quad Coarsening phase					& 1.707 	& 1.707 	& 1.707 	& 1.707 	& 1.707 	& 1.707\\
\quad Run phase					& 1.784 	& 1.780	& 1.776 	& 1.772	& 1.769	& 1.766\\
\\
VC--EXP4.P ($\gamma$)	 				& 1.765	& 1.764	& 1.762	& 1.761	& 1.759	& 1.758\\
\quad Coarsening phase					& 1.707 	& 1.707 	& 1.707 	& 1.707 	& 1.707 	& 1.707\\
\quad Run phase					& 1.777 	& 1.776	& 1.774	& 1.772	& 1.771	& 1.769\\
\\
VC--EXP4.P ($\eta$)					& 1.728	& 1.742	& 1.752	& 1.761	& 1.767	& 1.773\\
\quad Coarsening phase					& 1.707 	& 1.707 	& 1.707 	& 1.707 	& 1.707 	& 1.707\\
\quad Run phase					& 1.733 	& 1.749	& 1.762 	& 1.772	& 1.780	& 1.787\\
\hline
\end{tabularx}
\end{table}

The patterns that we observe in \cref{FIG:RATIO,FIG:BETA,FIG:GAMMA,FIG:ETA} mirror differences in the average welfare-level that the EXP4.P algorithm attains that are evident in \cref{TAB:RATIO,TAB:BETA}---experiments in which VC--EXP4.P converges slowly to the oracle rule correspond to those experiments in which it attains a lower average welfare-level.
\cref{THM:EXP4.P,THM:EXP4.P-VC,THM:EXP4.P-LES} and their proofs necessitate only fairly weak constraints on the values of the parameters of the EXP4.P algorithm with the definitions contained therein serving only, in practice, as recommendations.
\cref{THM:EXP4.P,THM:EXP4.P-VC,THM:EXP4.P-LES} and their proofs can, for instance, be suitably adjusted to facilitate more aggressive updating, such as \cref{FIG:RATIO,FIG:BETA,FIG:GAMMA,FIG:ETA,TAB:RATIO,TAB:BETA} suggest is necessary for the EXP4.P algorithm to attain a higher average welfare-level.
How much more aggressively the EXP4.P algorithm would need to update to attain an average welfare-level that is similar to that attained by $\tau$-EWM is, however, unclear but would appear substantial.
Not only does $\tau$-EWM outperform VC--EXP4.P across all of our experiments but does so whilst incurring far less computational expense.

\section{National Job Training Partnership Act Study}
\label{SEC:JTPA}
The National Job Training Partnership Act (JTPA) Study comprises 9,223 observations of subjects, their education level and prior earnings.
Applicants were enrolled onto the Study throughout the years 1987 to 1989, and were randomly allocated to one of two treatment arms---some applicants were extended training, job search assistance and other services provided by the JTPA, with the remaining applicants denied this support.  
Along with information collected prior to enrollment, the Study also collected administrative and survey data relating to applicants' earnings in the 30 months following its start.
Further details about the data and the Study can be found elsewhere (see, for instance, \citealp{BLOOM}). 
We restrict attention to a sample of 9,223 observations for which data on years of education and pre-programme earnings amongst the sample of adults (aged 22 years and older) used in the original evaluation of the programme and in subsequent studies (\citealp{ABADIE,BLOOM,HECKMAN-ICHIMURA-TODD}) is available. 
Applicants in the Study were assigned to the two treatment arms in the ratio of two to one.
Like \citet{KT}, we define the intervention to be the initial assignment of treatment, rather than the actual take-up due to the presence of non-compliance in the experiment.

\begin{figure}[t]
\centering
\includegraphics{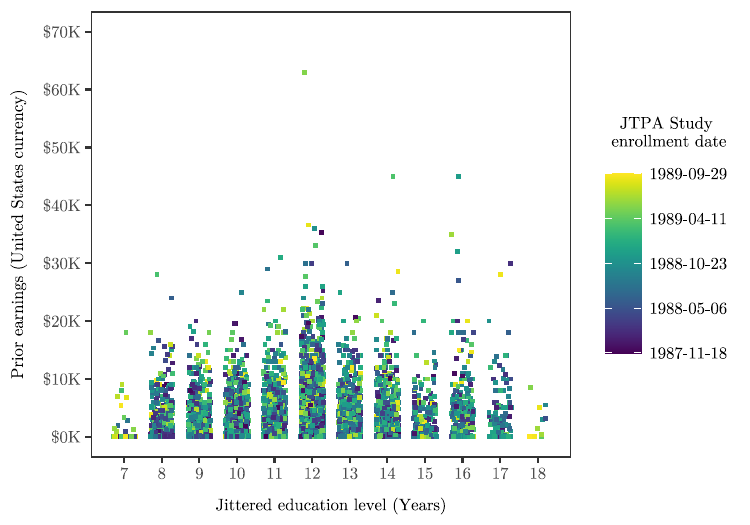}
\caption{Date of enrollment into the National JTPA Study.}
\label{FIG:ENROLLMENT}
\end{figure}

\subsection{Design}
We exploit the sequential arrival of applicants in the JTPA Study to test the performance of VC--EXP4.P.
We sort applicants by the date of their enrollment onto the Study, randomly ordering applicants with the same date of enrollment to mitigate any possible systematic issues arising from data entry---although it seems apparent from \cref{FIG:ENROLLMENT} that there are no systematic time trends in the distribution of observables during the Study.
We maintain this order throughout.
A restrictive feature of our analysis is that we assume that the social planner observes the realised outcome of an applicant \emph{before} the next applicant arrives. 
This is certainly an unrealistic assumption and presumes a favourable scenario.
The purpose of our analysis, however, is to study whether the EXP4.P algorithm performs well even in such an ideal situation. 
A more realistic analysis would replace the actual outcome by a short-run surrogate, as considered in \citet{Athey_Chetty_Imbens_Kang_2019}. 
We leave this extension for future research. 

The Study data reports applicants' earnings only for the treatment arm that they were allocated to---we do not observe their counterfactual earnings for obvious reasons. 
Although implementation of the EXP4.P algorithm does not require the social planner to observe the outcomes associated with other treatment arms, our simulation exercises do since the treatment implemented by the EXP4.P algorithm need not coincide with the treatment arm that applicants were allocated to.
We use random forest methods \citep{Breiman01} to regress applicants' earnings on their education level and prior earnings separately for each group of applicants---the applicants that were offered support, and the applicants that were denied support.
We subtract the average cost \$774 from applicants' earnings for those subjects that were offered support, prior to running the random forests.
Given that treatment was randomised in the JTPA study, we take these random forest regressions as estimators for the potential outcome regressions of $\mathrm{E}[y_{t}(1)| \mathbf{x}_{t}]$ and $\mathrm{E}[y(0)| \mathbf{x}_{t}]$.
We use these regressions to form a prediction about applicants' earnings for each of the two treatment arms--- with the applicant's education level and prior earnings as predictors---regardless of the treatment arm that they were allocated to.
We also use these regressions to construct an empirical distribution of residual earnings for each treatment arm, by subtracting the difference between an applicant's earnings and their predicted earnings for the treatment arm that they were allocated to.
By randomly drawing a residual earning from each empirical distribution and adding these to an applicant's predicted earnings for the corresponding treatment arm, we are able to construct a simulated outcome for each applicant for each of the two treatment arms, including the counterfactual one.
We repeat this process to generate many simulated samples and apply VC--EXP4.P (setting $\delta$ to the standard $95\%$-level, and setting the values of the tuning parameters as per \cref{THM:EXP4.P-LES}) to them to obtain a distribution of welfare-performance.  

As we do for our previous simulation exercises, we compare the performance of VC--EXP4.P with several benchmarks.
We implement (i.) the optimal rule, which we determine from the random forest;\footnote{%
The infeasible optimal rule allocates is to offer support whenever the random forest predicts that the difference between the potential outcomes plus the difference in the average residuals between the two treatment arms---which is not zero---is positive, and denies support otherwise.} 
(ii.) the infeasible quintic rule, which is an optimal assignment rule among the class of single indices formed by a fifth-order polynomial of the two covariates;
(iii.) an optimal assignment rule in the LES class;
(iv.) to treat everyone; 
(v.) to treat no-one; 
(vi.) $\tau$-EWM with the first 898 applicants used for coarsening $\mathfrak{F}$;\footnote{%
898 roughly corresponds to the number of applicants in the study that were enrolled during the first 90 days (i.e., one quarter).
For comparison, the suggested duration of the coarsening phase as per \cref{THM:EXP4.P-VC} is 609 applicants.} 
and (vii.) $\tau$-EWM over the first 609 applicants.
Given that VC--EXP4.P involves a coarsening phase, we can also compare the average welfare-level that is accrued during this sequence of observations to the average welfare-level during the subsequent run phase.

\begin{table}
\centering
\caption{Distributional features of the welfare-level attained by various estimators.}
\label{TAB:WELFARE}
\begin{tabularx}{\textwidth}{l*{8}C}
\hline
{Estimator}							& \multicolumn{7}{c}{{Quantile}}									& {Mean}\\
								\cmidrule{2-8}
								& 0\%		& 10\%	& 25\%	& 50\%	& 75\%	& 90\%	& 100\%	&\\
\hline\\
Infeasible optimal rule					& 16,821	& 17,144	& 17,238	& 17,338	& 17,433	& 17,529	& 17,876	& 17,337 \\
\\
Infeasible quintic rule					& 16,312	& 16,615	& 16,714	& 16,815	& 16,911	& 17,001	& 17,275	& 16,812\\
\\
Optimal LES rule						& 16,187	& 16,486	& 16,579	& 16,683	& 16,776	& 16,873	& 17,120	& 16,676\\
\\
Treat everyone						& 16,000	& 16,314	& 16,406	& 16,513	& 16,609	& 16,705	& 16,996	& 16,508\\
Treat no-one							& 14,865	& 15,122	& 15,218	& 15,319	& 15,414	& 15,493	& 15,861	& 15,315\\
\\
$\tau$-EWM with $\tau = 898$						& 15,007	& 15,598	& 15,776	& 16,003	& 16,220	& 16,393	& 16,841	& 16,000\\
$\tau$-EWM with $\tau = 609$			& 15,015	& 15,539	& 15,719	& 15,996	& 16,215	& 16,404	& 16,868	& 15,979\\
\\
VC--EXP4.P 							& 15,430	& 15,726	& 15,822	& 15,924	& 16,020	& 16,117	& 16,366	& 15,922\\
\quad Coarsening phase					& 14,172	& 15,193	& 15,555	& 15,971	& 16,399	& 16,746	& 17,814	& 15,967\\
\quad Run phase					& 15,395	& 15,711	& 15,817	& 15,918	& 16,018	& 16,122	& 16,363	& 15,918\\				
\hline
\end{tabularx}
\end{table}

\subsection{Implementation}
Aside from enrollment date, \cref{FIG:ENROLLMENT} also reports education level and prior earnings in the sample. 
We note two things about the distribution of covariates.  
First, education level is recorded on a discrete scale (although data is artificially jittered for legibility), in years.
An implication is that any assignment rule that allocates an applicant to a particular treatment arm must also allocate applicants with the same education level and lower prior earnings (or higher, depending upon the \emph{sign} of the rule) to the same treatment arm.
Second, prior earnings are highly concentrated for every education level. 
In particular, many applicants have zero prior earnings.
An implication is that the Study can, in practice, be reduced to a far smaller number of unique observations (of education level and prior earnings) for the purpose of refining the assignment rules.\footnote{%
The 609 applicants that are considered during the coarsening phase reduce to 365 unique observations.}

To ensure that the potential outcomes are non-negative, we add $\$34,027$ to every observation, subsequently subtracting this amount before reporting our results.
Specifically, we add $\$34,027$ so as to ensure \cref{ASS:UNIFORM}, since applicants' simulated earnings can otherwise be negative due to the lack of restriction that we impose on the random forests (predicted earnings and residual earnings can both be negative).
We set $M$ to $\$257,300$, which is the sum of the maxima of the predicted earnings and residual earnings that are returned by the random forests following this normalisation.\footnote{%
We note that the maximum potential outcome that could be observed in each simulation during the first 609 periods ranges from as little as $\$98,023$ to as much as $\$198,337$; 
the standard deviation of these observations is less than $\$20,000$ in each simulation.
How to set $M$ in the absence of full information is an open question.}

To reduce the LES class, we utilise the incremental enumeration algorithm described in \Cref{SEC:HYPERPLANES}.
We find that 104,252 assignment rules are compatible with the education level and prior earnings of the first 609 applicants to enroll on the Study.
For comparison, \citet{EF} suggests that up to 370,274 assignment rules could be admitted if the Study data did not include duplicates or applicants with the same prior earnings (particularly, zero prior earnings) and different education levels, or if education level were not discrete.
As such, the EXP4.P algorithm can incur substantial computational expense---especially in terms of memory usage.

\begin{figure}[t]
\centering
\includegraphics{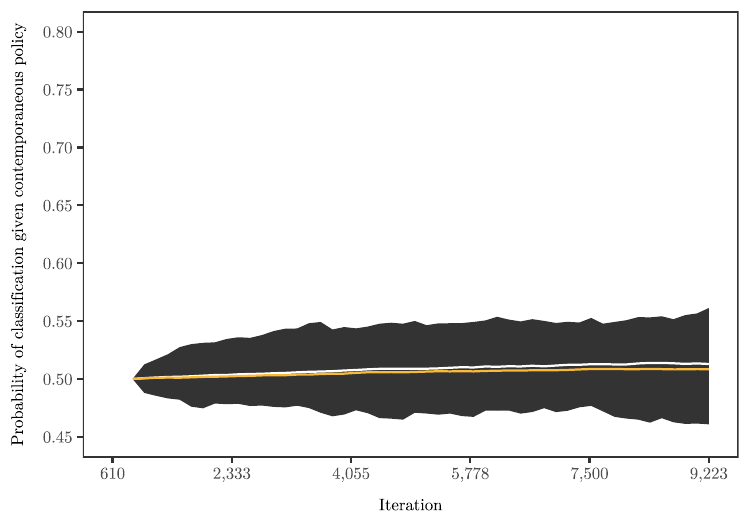}
\caption{The simulated performance of VC--EXP4.P based on the samples in the National JTPA Study.}
\label{FIG:MISCLASSIFICATION}
\end{figure}

\subsection{Results}
We construct 1,000 simulated samples that we use to assess the performance of VC--EXP4.P and our benchmark procedures.
We report quantiles and the mean of the welfare-level that each method attains in \cref{TAB:WELFARE}.
We plot the probability of classification as \cref{FIG:MISCLASSIFICATION}.
The black ribbon captures the probabilities that the policy enacts the same intervention as the optimal LES rule at every iteration in each of the 1,000 artificial datasets that we construct;
the white line that bisects the ribbon is the median probability amongst these simulated datasets;
the orange line that dissects the ribbon is the corresponding median probability when the comparison is the intervention dictated by the infeasible oracle rule---rather than the optimal LES rule.

\begin{figure}[t]
\centering
\includegraphics{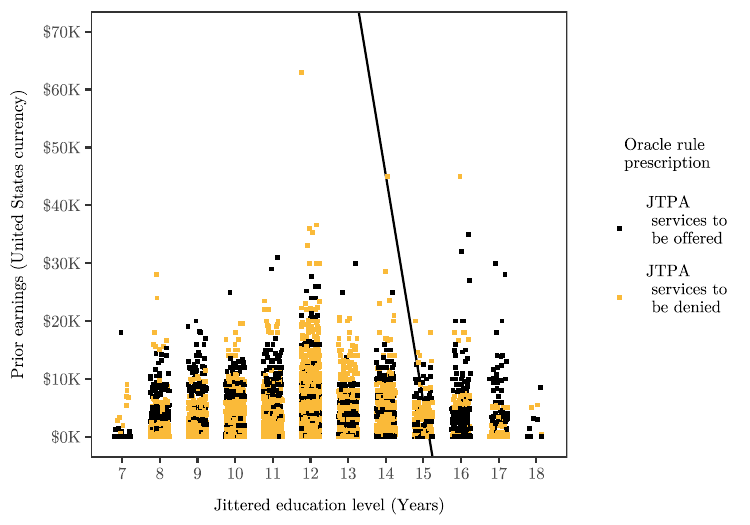}
\caption{The allocation that is prescribed by the infeasible oracle rule for the National JTPA Study.}
\label{FIG:ORACLE}
\end{figure}

We plot the allocation that is prescribed for the Study as \cref{FIG:ORACLE}.
We contrast the prescription of the infeasible optimal rule with the prescription of the optimal LES rule that is to offer support to those subjects to the left of the downwards-sloping line.
We note several findings.

First, on average, applicants benefit from training, job search assistance and other services provided by the JTPA, as is illustrated by the higher welfare-level attained by treating everyone versus treating no-one.
The magnitude and sign of this effect is, however, known to be heterogeneous;
individualised assignment can be beneficial, as is shown in \citet{KT}. 
Comparison of the welfare-level attained by the infeasible optimal rule (the plug-in rule based on the random forest estimates) and by the treat everyone policy are in line with heterogeneity in the sign of the treatment effect. 
The difference in welfare-level shrinks, however, if we compare the results for the optimal quintic rule or optimal LES rule with the treat everyone policy.

Second, in terms of the average welfare for in-sample subjects, neither the EXP4.P algorithm nor EWM performs well, as is evidenced by the inferior mean welfare-levels that are attained by VC--EXP4.P and $\tau$-EWM versus the treat everyone policy. 
\cref{THM:EXP4.P-VC} suggests that the welfare-level of VC-EXP4.P converges to the welfare-level of the optimal LES rule, but we do not observe this happening.  
This result is also indicative of the Study having a high degree of difficulty. 
The counterpart to \cref{EQ:DIFFICULTY} that we employ here is the probability that the sign of the difference in potential outcomes---as predicted by the random forest or by the closest assignment rule in the LES class---is different from the sign of the difference in potential outcomes that is realised.
We find that the Study has a difficulty rating of 44.0\% and 47.6\% (relative to a maximum of 50.0\%) according to these measures, respectively. 
This conclusion is also supported by \cref{FIG:MISCLASSIFICATION}, which shows that the probability of correct classification does not grow over time and remains close to one half---the same as can be achieved by pure randomisation using an unbiased coin. 
The infeasible optimal rule is highly non-linear, and rules in the LES class are unable to effectively replicate it---resulting in the large welfare gap between the infeasible optimal rule and the optimal LES rule.

Third, VC--EXP4.P attains a similar welfare-level to $\tau$-EWM.
This indicates that, in the JTPA Study sample, the welfare gain from performing sequential learning using VC--EXP4.P is small compared with the simpler two-stage learning procedure of $\tau$-EWM, despite a substantial difference in computational cost.

In summary, the results of our simulations---based upon the JTPA Study sample---indicate limited benefits to implementing the EXP4.P algorithm, even in an unrealistically ideal scenario in which the social planner can update the assignment policies of as many as 8,000 subjects and in which treatment response can be observed immediately after assignment.
We attribute this limited benefit to substantial heterogeneity and non-linearity in treatment effect in the JTPA Study sample (i.e., the JTPA Study sample is a difficult sample).
Although the effect of non-linearity can be mitigated somewhat by increasing the VC dimension of $\mathfrak{F}$---evidenced by our results for the infeasible quintic rule---increasing the complexity of $\mathfrak{F}$ leads to a substantial increase in the computational cost of implementing the EXP4.P algorithm, particularly during the coarsening phase of VC--EXP4.P.

\section{Conclusions}
This paper studies applicability of the existing EXP4.P algorithm to the problem of treatment choice, where the social planner's goal is to learn an optimal individualised treatment assignment policy within a constrained class of assignment rules.
For general bounded outcome variables, we provide a welfare-regret guarantee for the EXP4.P algorithm, and a practical method for coarsening the class of assignment rules---a necessary step when this class is infinite with complexity controlled by a finite VC dimension---that exploits the similarity of the coarsening problem to a hyperplane arrangement and cell enumeration problem. 

We study the suitability of the EXP4.P algorithm to treatment choice problems through numerical analysis.
Using a novel simulation design that is based upon the JTPA Study sample and that mimics the empirical context of assignment to job training, we assess the performance of the EXP4.P algorithm relative to oracle assignment rules and a non-adaptive benchmark policy based upon static EWM.
Our main finding is somewhat discouraging.
Specifically, we find that the EXP4.P algorithm can perform poorly in situations in which the standard deviation of unobserved heterogeneity is large relative to the size of conditional average treatment effects or in which non-linearity of these effects limits the welfare gain of simple (linear) assignment policies.\footnote{%
We conduct further experiments as to the effect of non-linearity of conditional average treatment effects on the performance of the EXP4.P algorithm supporting this statement, the results of which are available upon request.}
The JTPA Study sample exhibits these properties and, as a result, implementing the EXP4.P algorithm does not deliver a notable welfare gain relative to the policy recommended by non-sequential EWM.
Moreover, although our theoretical results establish uniform convergence of welfare-regret, we do not observe this convergence in practice.

We regard several issues as open questions and interesting topics for future research. 
First, implementation of the EXP4.P algorithm, importantly, requires that the time horizon (i.e., the number of periods over which allocation is to occur) is fixed and known in advance.
This assumption is restrictive.
Second, our analysis of the EXP4.P algorithm when the class of assignment rules is infinite with complexity controlled by a finite VC dimension is limited to the case of two treatment arms.
Extending this analysis to more treatment arms complicates the coarsening step that must be undertaken, and the link to hyperplane arrangements needs to be properly modified.
Whether information about treatment response yielded during the coarsening step can be incorporated (it is currently discarded) to improve the performance of the EXP4.P algorithm is unclear, and is pertinent even in the case of two treatment arms.
Finally, our analysis of the EXP4.P algorithm does not consider budget or capacity constraints.
Such constraints are important in many public policy applications. 

\begin{appendix}
\section{Proofs}
\label{SEC:PROOFS}
\begin{proof}[Proof of \cref{THM:EXP4.P}]
Our proof closely follows the proof given in \cite{EXP4.P} except that our proof explicitly introduces the upper bound of the outcome variable $M$. Throughout, we subscript the expectation operator by time, to denote the conditional expectation that is contingent upon the information set possessed by the social planner in a given period;
we subscript the expectation operator by $i$ and time, to denote the probability distribution over treatment arms that is induced by $f^{i}$ in a given period.
We otherwise maintain the notation defined in the main text, additionally defining $s_{t}^{i}=\sum_{k=1}^{K} y_{t}(k) \cdot f_{k}^{i}(\mathbf{x}_{t})$ and $S_{t}^{i}=\sum_{\ell=1}^{t}s_{\ell}^{i}$. 
The provisos of the theorem dictate that $\mathfrak{F}$ includes a randomising expert (ascribing equal probability to each treatment arm), and $T\geq\omega^{2}\vee [4 K\cdot\ln(N)]$.
These requirements translate directly, to $0\leq\beta\leq 1/M$ and $0\leq\gamma<1/2$.

First, we establish a large deviation inequality for $\tilde{S}_{T}^{i}$ relative to $S_{T}^{i}$ and, in particular, for a best performing. Let $\mathrm{E}_{t}$ denote the conditional expectation given $(\mathcal{I}_t, \mathbf{x}_t)$ and consider
\begin{align}
&\mathrm{E}_{t}(\exp(\beta\cdot s^{i}_{t}-\beta\cdot\tilde{s}_{t}^{i}))\notag\\
&\quad=\label{EQ:JENSEN-A}
\mathrm{E}_{t} \left( \exp \left( \textstyle\sum_{k=1}^{K}(\beta\cdot y_{t}(k)-\beta\cdot\tilde{y}_{t}(k)) f_{k}^{i}(\mathbf{x}_{t}) \right) \right)
\\
&\quad\leq\label{EQ:JENSEN-B}
\mathrm{E}_{t}\left( \textstyle\sum_{k=1}^{K}\exp(\beta\cdot y_{t}(k)-\beta\cdot\tilde{y}_{t}(k)) \cdot f_{k}^{i}(\mathbf{x}_{t}) \right)
\\
&\quad=\label{EQ:JENSEN-C}
\textstyle\sum_{k=1}^{K}  \mathrm{E}_{t}(\exp(\beta\cdot y_{t}(k)-\beta\cdot[\beta\cdot M^{2}+y_{t}(k)\cdot 1(k_{t}=k)]/p_{t}(k))) \cdot f_{k}^{i}(\mathbf{x}_{t}),
\end{align}
where \cref{EQ:JENSEN-B} makes use of Jensen's Inequality.
Since $\beta\cdot y_{t}(k)\leq 1$ by \cref{ASS:UNIFORM} together with our requirement upon the magnitude of $\beta$ as assumed in \cref{THM:EXP4.P}, an exponential bound $\exp(z)\leq 1+z+z^{2}$ whenever $z\leq 1$ implies
\begin{align}
&\mathrm{E}_{t}(\exp(\beta\cdot y_{t}(k)-\beta\cdot\tilde{y}_{t}(k)))\notag\\
&\quad=\label{EQ:TRICK-A}
\mathrm{E}_{t}(\exp(\beta\cdot y_{t}(k)-\beta\cdot y_{t}(k)\cdot 1(k_{t}=k)/p_{t}(k)))\cdot\exp(p_{t}(k)/[\beta\cdot M]^{2})
\\
&\quad\leq\label{EQ:TRICK-B}
\mathrm{E}_{t}(1+[\beta\cdot y_{t}(k)-\beta\cdot y_{t}(k)\cdot 1(k_{t}=k)/p_{t}(k)]^{2})\cdot\exp(p_{t}(k)/[\beta\cdot M]^{2})
\\
&\quad\leq\label{EQ:TRICK-C}
[1+\beta^{2}\cdot\mathrm{E}_{t}(y_{t}(k)^{2})/p_{t}(k)]\cdot\exp(p_{t}(k)/[\beta\cdot M]^{2})
\\
&\quad\leq\label{EQ:TRICK-D}
1.
\end{align}
We note that \cref{EQ:TRICK-A} to \cref{EQ:TRICK-B} relies on the fact that the realised treatment is drawn according to $\mathbf{p}_{t}$, with the expectation of the linear term in the exponential bound then equals to zero,
and \cref{EQ:TRICK-C} to \cref{EQ:TRICK-D} makes use of the alternative exponential bound that is $1+z\leq\exp(z)$, and \cref{ASS:UNIFORM} together with our requirement upon the magnitude of $\beta$ being bounded by $1/M$ as implied by the assumptions of \cref{THM:EXP4.P}.
It then follows that
\begin{equation}
\mathrm{E}_{P}(\exp(\beta\cdot S_{T}^{i}-\beta\cdot\tilde{S}_{T}^{i}))
\leq\label{EQ:LEQ-1}
\textstyle\prod_{t=1}^{T}\mathrm{E}_{t}(\exp(\beta\cdot s^{i}_{t}-\beta\cdot\tilde{s}_{t}^{i}))
\leq 1.
\end{equation}
By Markov's inequality and \cref{EQ:LEQ-1}, we obtain
\begin{align}
&P(S^{i}_{T}-\tilde{S}^{i}_{T}\geq[\beta\cdot K\cdot M^{2}]\cdot T)\notag\\
&\quad=\label{EQ:MARKOV-A}
P(\exp(\beta\cdot S^{i}_{T}-\beta\cdot\tilde{S}^{i}_{T})\geq\exp(\beta\cdot[\beta\cdot K\cdot M^{2}]\cdot T))
\\
&\quad\leq\label{EQ:MARKOV-B}
\mathrm{E}_{P}(\exp(\beta\cdot S^{i}_{T}-\beta\cdot\tilde{S}^{i}_{T}))\cdot\exp(-\beta\cdot[\beta\cdot K\cdot M^{2}]\cdot T)
\\
&\quad\leq\label{EQ:MARKOV-C}
\exp(-\beta \cdot A \cdot T ),
\end{align}
where $A = \beta K M^2$. By the choice of $\beta$ as per \cref{THM:EXP4.P}, we obtain
\begin{equation}
P(S_{T}^{i}-\tilde{S}^{i}_{T}\geq A \cdot T)
\leq\label{EQ:ANY BOUND}
\delta/N
\end{equation}
Note that \cref{EQ:ANY BOUND} holds for an arbitrary expert, building on which we can obtain a large deviation inequality of $\max_i(\tilde{S}_{T}^i)$ relative to $\max_i(S_{T}^i)$. Let $i^{\ast} \doteq \argmax_i S_{T}^i$ and, note that
\begin{align*}
& \left\{ \max_i (S_{T}^i) \geq \max_i (\tilde{S}_{T}^i) + A \cdot T \right\}  \subset \left\{ S_{T}^{i^{\ast}} \geq (\tilde{S}_{T}^{i^{\ast}}) + A \cdot T  \right\} \\
&  \subset \{ \exists i \mspace{10mu} S_{T}^i \geq \tilde{S}_{T}^i + A \cdot T \} = \bigcup_{i=1}^N \{ S_{T}^i \geq \tilde{S}_{T}^i + A \cdot T \}.
\end{align*}
Hence, by applying the union bound, we obtain 
\begin{align}
P(\max_i (S_{T}^{i})-A\cdot T\geq\textstyle\max_{i}(\tilde{S}_{T}^{i}))
\leq\label{EQ:LARGE DEVIATION}
\textstyle\sum_{i=1}^{N}P(S_{T}^{i}-\tilde{S}_{T}^{i}\geq A \cdot T)
\leq\delta
\end{align}

Next, we establish a high probability bound for the welfare regret using \cref{EQ:LARGE DEVIATION}. Define 
\begin{equation}
C_{t}
\doteq\label{EQ:DENOMINATOR}
\begin{cases} N \mspace{10mu}  & \text{for $t=0$} \\  \textstyle\sum_{i=1}^{N}\exp(\eta\cdot\tilde{S}_{t}^{i}) \mspace{10mu}  & \text{for $t=1, \dots, T$}
\end{cases}
\end{equation}
which we can relate to a (prospective) policy weight using $q_{t}^{i}=\exp(\eta\cdot\tilde{S}_{t-1}^{i})/C_{t-1}$, where we interpret $\tilde{S}^i_0 = 0$ for $i=1, \dots, N$. 
Hence,
\begin{align}
C_{t}/C_{t-1}
&=\label{EQ:TELESCOPING-A}
\textstyle\sum_{i=1}^{N}\exp(\eta\cdot\tilde{s}^{i}_{t})\cdot\exp(\eta\cdot\tilde{S}^{i}_{t-1})/C_{t-1}
\\
&=\label{EQ:TELESCOPING-B}
\textstyle\sum_{i=1}^{N}\exp(\eta\cdot\tilde{s}^{i}_{t})\cdot q_{t}^{i}
\\
&\leq\label{EQ:TELESCOPING-C}
1+\eta\cdot\underbracket{\textstyle\sum_{i=1}^{N}\tilde{s}^{i}_{t}\cdot q_{t}^{i}}_{B_{1t}}+\eta^{2}\cdot\underbracket{\textstyle\sum_{i=1}^{N}[\tilde{s}^{i}_{t}]^{2}\cdot q_{t}^{i}}_{B_{2t}}.
\end{align}
We note that \cref{EQ:TELESCOPING-B} to \cref{EQ:TELESCOPING-C} relies on the exponential bound previously used in \cref{EQ:TRICK-B}, which is applicable here by \cref{ASS:UNIFORM} together with the given specifications of $\beta$, $\gamma$ and $\eta$ in the current theorem.
The two labelled terms in \cref{EQ:TELESCOPING-C} can then be bounded as follows.
\begin{align}
B_{1t}
&=\label{EQ:A TERM-B}
\textstyle\sum_{i=1}^{N}\textstyle\sum_{k=1}^{K}\tilde{y}_{t}(k)\cdot f_{k}^{i}(\mathbf{x}_{t})\cdot q_{t}^{i}
\\
&=\label{EQ:A TERM-C}
\textstyle\sum_{k=1}^{K}[\beta\cdot M^{2}+y_{t}(k_{t})\cdot 1(k_{t}=k)]/p_{t}(k)\cdot\textstyle\sum_{i=1}^{N}f_{k}^{i}(\mathbf{x}_{t})\cdot q_{t}^{i}
\\
&\leq\label{EQ:A TERM-D}
1/[1-\gamma]\cdot\textstyle\sum_{k=1}^{K}[\beta\cdot M^{2}+y_{t}(k_{t})\cdot 1(k_{t}=k)]
\\
&=\label{EQ:A TERM-E}
\left( A + y_t(k_t) \right)/[1-\gamma]
\end{align}
We note that \cref{EQ:A TERM-C} to \cref{EQ:A TERM-D} is by the construction of $\mathbf{p}_{t}$.\footnote{%
Specifically, $p_{t}(k)=[1-\gamma]\cdot\textstyle\sum_{i=1}^{N}f_{k}^{i}(\mathbf{x}_{t})\cdot q_{t}^{i}+\gamma/K$, or $p_{t}(k)\geq[1-\gamma]\cdot\textstyle\sum_{i=1}^{N}f_{k}^{i}(\mathbf{x}_{t})\cdot q_{t}^{i}$.
Similarly, $p_{t}(k)\geq\gamma/K$.}
\begin{align}
B_{2t}
&=\label{EQ:B TERM-A}
\textstyle\sum_{i=1}^{N}[\tilde{\mathbf{y}}_{t}^{\intercal}f^{i}(\mathbf{x}_{t})]^{2}\cdot q_{t}^{i}
\\
&\leq\label{EQ:B TERM-B}
\textstyle\sum_{i=1}^{N}\textstyle\sum_{k=1}^{K}\tilde{y}_{t}(k)^{2}\cdot f_{k}^{i}(\mathbf{x}_{t})\cdot q_{t}^{i}
\\
&=\label{EQ:B TERM-C}
\textstyle\sum_{k=1}^{K}[[\beta\cdot M^{2}+y_{t}(k_t)\cdot 1(k_{t}=k)]/p_{t}(k)]^{2}\cdot\textstyle\sum_{i=1}^{N}f_{k}^{i}(\mathbf{x}_{t})\cdot q_{t}^{i}
\\
&\leq\label{EQ:B TERM-D}
2\textstyle\sum_{k=1}^{K}[[\beta\cdot M^{2}]^{2}+y_{t}(k_t)^{2}\cdot 1(k_{t}=k)]/p_{t}(k)\cdot\textstyle\sum_{i=1}^{N}f_{k}^{i}(\mathbf{x}_{t})\cdot q_{t}^{i}/p_{t}(k)
\\
&\leq\label{EQ:B TERM-E}
2/[1-\gamma]\cdot\textstyle\sum_{k=1}^{K}[[\beta\cdot M^{2}]^{2}+y_{t}(k_{t})^{2}\cdot 1(k_{t}=k)]/p_{t}(k)
\\
&\leq\label{EQ:B TERM-F}
2/(1-\gamma) \cdot \left[ A^2/\gamma + M\cdot y_{t}(k_{t})/p_{t}(k_{t}) \right]
\end{align}
We note that \cref{EQ:B TERM-A} to \cref{EQ:B TERM-B} makes use of the Cauchy--Schwarz Inequality and the fact that $f$ is constrained to the unit simplex;\footnote{%
We use the expectation version of the Cauchy--Schwarz Inequality here, with $f^{i}$ playing the role of probability weights.}
\cref{EQ:B TERM-C} to \cref{EQ:B TERM-D} makes use of the quadratic inequality $[a+b]^{2}\leq 2[a^{2}+b^{2}]$;
and \cref{EQ:B TERM-D} to \cref{EQ:B TERM-E} and \cref{EQ:B TERM-E} to \cref{EQ:B TERM-F} are by the construction of $\mathbf{p}_{t}$ and \cref{ASS:UNIFORM}.
Applying a logarithmic telescope to \cref{EQ:TELESCOPING-A}, we obtain
\begin{align}
&\ln(C_{T}/C_{0})\notag\\
&\quad=\label{EQ:LOG TELESCOPE-A}
\textstyle\sum_{t=1}^{T}\ln(C_{t}/C_{t-1})
\\
&\quad\leq\label{EQ:LOG TELESCOPE-B}
\textstyle\sum_{t=1}^{T}\ln(1+\eta\cdot B_{1t} + \eta^{2}\cdot B_{2t} )
\\
&\quad\leq\label{EQ:LOG TELESCOPE-C}
\textstyle\sum_{t=1}^{T}[\eta\cdot B_{1t} + \eta^{2} \cdot B_{2t} ]
\\
&\quad\leq\label{EQ:LOG TELESCOPE-D}
\eta\cdot T\cdot[1+2 \eta \cdot A  / \gamma]\cdot A/[1-\gamma]+\eta \cdot \textstyle\sum_{t=1}^{T}[1+2\eta\cdot M /p_{t}(k_{t})]\cdot y_{t}(k_{t})/[1-\gamma]
\\
&\quad=\label{EQ:LOG TELESCOPE-E}
\eta/[1-\gamma]\cdot[\underbracket{[1+\beta\cdot M] \cdot A \cdot T}_{A^{\prime}}+\textstyle\sum_{t=1}^{T}y_{t}(k_{t})+\gamma/K\cdot \textstyle\sum_{t=1}^{T}y_{t}(k_{t})/p_{t}(k_{t})]
\end{align} 
We note that \cref{EQ:LOG TELESCOPE-A} to \cref{EQ:LOG TELESCOPE-B} implements \cref{EQ:TELESCOPING-C};
we note that \cref{EQ:LOG TELESCOPE-B} to \cref{EQ:LOG TELESCOPE-C} makes use of the logarithmic inequality $\ln(1+z)\leq z$;
we note that \cref{EQ:LOG TELESCOPE-C} to \cref{EQ:LOG TELESCOPE-D} implements \cref{EQ:A TERM-E,EQ:B TERM-F};
and we note that \cref{EQ:LOG TELESCOPE-D} to \cref{EQ:LOG TELESCOPE-E} is by rearrangement and by the specifications of $\beta$, $\gamma$ and $\eta$ as per \cref{THM:EXP4.P}.
Rewriting \cref{EQ:LOG TELESCOPE-E} via multiplication by the inverse of the common term, we obtain the following upper bound.
\begin{align}
&[1-\gamma]/\eta\cdot\ln(C_{T}/C_{0})\notag\\
&\quad\leq\label{EQ:UB-A}
A^{\prime} +\textstyle\sum_{t=1}^{T}y_{t}(k_{t})+\gamma/K\cdot\textstyle\sum_{t=1}^{T}y_{t}(k_{t})/p_{t}(k_{t})
\\
&\quad\leq\label{EQ:UB-B}
A^{\prime} +\textstyle\sum_{t=1}^{T}y_{t}(k_{t})+\gamma\cdot\textstyle\sum_{t=1}^{T}\textstyle\sum_{k=1}^{K}\tilde{y}_{t}(k)/K
\\
&\quad=\label{EQ:UB-C}
A^{\prime} +\textstyle\sum_{t=1}^{T}y_{t}(k_{t})+\gamma\cdot\tilde{S}_{T}^{\textrm{random}}
\\
&\quad\leq\label{EQ:UB-D}
A^{\prime} +\textstyle\sum_{t=1}^{T}y_{t}(k_{t})+\gamma\cdot\textstyle\max_{i}(\tilde{S}_{T}^{i}),
\end{align}
where $\tilde{S}_{T}^{\textrm{random}}$ is the cumulative score of the pure randomizing expert that enacts each possible action with equal probability. 

We also obtain the following lower bound.
\begin{align}
&[1-\gamma]/\eta\cdot\ln(C_{T}/C_{0})\notag\\
&\quad\geq\label{EQ:LB-A}
[1-\gamma]\cdot\textstyle\max_{i}(\tilde{S}_{T}^{i})-[1-\gamma]/\eta\cdot\ln(C_{0})
\\
&\quad\geq\label{EQ:LB-B}
[1-\gamma]\cdot\textstyle\max_{i}(\tilde{S}_{T}^{i})-[1-\gamma]/\eta\cdot\ln(N),
\end{align}
where \cref{EQ:LB-A} follows since $\textstyle\sum_{i=1}^N \exp(\eta \tilde{S}_{T}^i) \geq \exp(\eta \max_{i} (\tilde{S}_{T}^i)) $ and \cref{EQ:LB-B} follows by the uniformity of the initial weights as specified in \cref{THM:EXP4.P}.
Combining \cref{EQ:UB-D,EQ:LB-B} yields 
\begin{equation}
\label{EQ:WORKING INEQUALITY}
[1-2\gamma]\cdot\textstyle\max_{i}(\tilde{S}_{T}^{i})-\textstyle\sum_{t=1}^{T}y_{t}(k_{t})
\leq
A^{\prime} + \underbracket{[1-\gamma]/\eta\cdot\ln(N)}_{A''}.
\end{equation}
Each of the two terms in the right-hand side of \cref{EQ:WORKING INEQUALITY} can be bounded as follows:
\begin{equation}
\label{EQ:PART CIRCLED C}
A'
\leq
2 \cdot A\cdot T
=
2M\cdot\sqrt{K\cdot T\cdot\ln(N/\delta)},
\end{equation}
where this inequality follows by $\beta \leq 1/M$ as implied by the assumptions of \cref{THM:EXP4.P}.
\begin{equation}
\label{EQ:PART CIRCLED D}
A''
=
2M\cdot\sqrt{K\cdot\ln(N)}\cdot[\sqrt{T}-\sqrt{K\cdot\ln(N)}]
\leq
2M\cdot\sqrt{K\cdot T\cdot\ln(N/\delta)},
\end{equation}
where we note that $0<\delta<1$.
Applying \cref{EQ:LARGE DEVIATION}---more precisely, its complement---to the left-hand side of \cref{EQ:WORKING INEQUALITY} and rearranging, we determine that 
\begin{equation}
\hat{R}_{T}
=
\max_i (S_T^i ) - \textstyle\sum_{t=1}^{T}y_{t}(k_{t})
\leq
2\gamma\cdot \max_i S_{T}^i+[1-2\gamma]\cdot A\cdot T+ A' + A''
\end{equation}
holds with probability at least $1-\delta$.
Utilising the results of \cref{EQ:PART CIRCLED C,EQ:PART CIRCLED D}, making use of the support condition $ \max_i ( S^{i}_{T} ) \leq M\cdot T$, and noting $(1-2\gamma) \cdot A \cdot T \leq M \sqrt{K \cdot T \cdot \ln (N / \delta)}$, we obtain the conclusion of \cref{THM:EXP4.P}.
\end{proof}

\begin{proof}[Proof of \cref{THM:EXP4.P-VC}]
Let $f^{\ast} \in \argsup_{f \in \mathfrak{F}} W_T(f)$, and $g^{\ast} \in \mathfrak{G} $ be a best expert in the run phase in the sense of $g^{\ast} \in \argmax_{g \in \mathfrak{G}} \sum_{t= \lceil \tau \rceil + 1}^T g(\mathbf{x}_t)^{\intercal} \mathbf{y}_t$. 
For the class of experts with non-randomized assignments, Sauer's lemma bounds the cardinality of $ \{ (f(x_1),\dots,f(x_{T'})) : f \in \mathcal{F} \}$  from above by $\left( \frac{eT'}{D} \right)^D$. Consider the following decomposition of $R_T$:
\begin{align}
R_T & = \sup_{f \in \mathfrak{F}} E_P \left( \textstyle\sum_{t=1}^T f(\mathbf{x}_t)^{\intercal} \mathbf{y}_t \right) - \textstyle\sum_{t=1}^T y_t(k_t) \notag \\
& = \underbracket{\lceil \tau \rceil E_P(f^{\ast}(\mathbf{x}_t)^{\intercal} \mathbf{y}_t)}_{\text{(i)}} - \underbracket{\left[  \textstyle\sum_{t= \lceil \tau \rceil + 1}^T f^{\ast}(\mathbf{x}_t)^{\intercal} \mathbf{y}_t  - E_P \left( \textstyle\sum_{t= \lceil \tau \rceil + 1}^T f^{\ast}(\mathbf{x}_t)^{\intercal} \mathbf{y}_t \right) \right] }_{\text{(ii)}}  \label{EQ:RT_decomposition} \\
& \mspace{20mu} + \underbracket{ \textstyle\sum_{t = \lceil \tau \rceil + 1}^{T}  ( f^{\ast}(\mathbf{x}_t) - g^{\ast}(\mathbf{x}_t) )^{\intercal} \mathbf{y}_t  }_{\text{(iii)}} + \underbracket{ \textstyle\sum_{t= \lceil \tau \rceil + 1}^{T} \left( g^{\ast}(\mathbf{x}_t)^{\intercal} \mathbf{y}_t - y_t(k_t) \right) }_{\text{(iv)}}.
\end{align} 
Here, term (i) corresponds to the error accumulated during the coarsening phase for the first $\lceil\tau\rceil$ periods.
The sum of terms (ii) and (iii) capture the error accumulated by using $\mathfrak{G}$ instead of $\mathfrak{F}$.
Term (iv) corresponds to the regret associated with the run phase when the EXP4.P--F variant is implemented for the next $T-\lceil\tau\rceil$ periods with the coarsened class $\mathfrak{G}$.

We bound each of (i) - (iv) terms separately. By \cref{ASS:UNIFORM}, (i) can be upper bounded by $\lceil \tau \rceil M$. To bound (ii), we invoke Chernoff's inequality, 
\begin{equation} \label{EQ:Chernoff}
P \left(  \textstyle\sum_{t= \lceil \tau \rceil + 1}^T f^{\ast}(\mathbf{x}_t)^{\intercal} \mathbf{y}_t  - E_P \left( \textstyle\sum_{t= \lceil \tau \rceil + 1}^T f^{\ast}(\mathbf{x}_t)^{\intercal} \mathbf{y}_t \right) \leq M \sqrt{(T - \lceil \tau \rceil)/2 \cdot \ln(3/\delta)}  \right) \geq 1 - \delta/3. 
\end{equation}
Note that (iv) corresponds to the empirical regret of the EXP4.P-F algorithm for $T - \lceil \tau \rceil$ periods under the coarsen class $\mathfrak{G} $ whose cardinality is bounded by $|\mathfrak{G}| \leq \left( \frac{e \lceil \tau \rceil}{D} \right)^D$  by Sauer's lemma. 
Hence, with probability at least $1- \delta/3$, it holds
\begin{equation} \label{EQ:Term_iv_bound}
\text{(iv)} \leq c_1 M \cdot\sqrt{[T-\lceil\tau\rceil]\cdot[2D\cdot\ln(\lceil\tau\rceil\cdot e/D) + \ln(3/\delta)]},
\end{equation}
where $c_1$ is a universal constant.

Given a sequence of subjects, note that (iii) can be bounded by $M \cdot \| f^{\ast} - g^{\ast} \|$, where  
\begin{equation*}
\| f^{\ast} - g^{\ast} \| \doteq \textstyle\sum_{t=\lceil\tau\rceil+1}^{T}1(f^{\ast}(\mathbf{x}_{t})\neq g^{\ast}(\mathbf{x}_{t})). 
\end{equation*}
 Under \cref{ASS:STATIONARY}, this sequence is as likely to occur as any other sequence of subjects, of which there are a total of $T!$ such permutations.
Denote a permutation by a one-to-one function $\chi : \{1, \dots, T \} \to \{1, \dots, T \}$, and consider a uniform distribution over the $T!$ permutations denoted by $P_{\chi}$.
Conditional on a given sequence and the covariates associated with it, we have 
\begin{align}
&P_{\chi}(\textstyle\sum_{t=\lceil\tau\rceil+1}^{T}1(f^{\ast}(\mathbf{x}_{\chi(t)})\neq g^{\ast}(\mathbf{x}_{\chi(t)}))>m)
\notag\\
&\qquad\leq\label{EQ:PROB-A}
P_{\chi}(\exists f\in\mathfrak{F}\text{ and }h\in\mathfrak{F}\text{ satisfying } \|f-h\|>m\text{ and }f(\mathbf{x}_{\chi(t)})=h(\mathbf{x}_{\chi(t)})\text{ for }t\leq\lceil\tau\rceil)
\\
&\qquad\leq\label{EQ:PROB-B}
\left|\mathfrak{F}\right|^{2}\cdot\textstyle\binom{T-m}{\lceil\tau\rceil}/T! \\
&\qquad\leq\label{EQ:PROB-C}
[e\cdot T/D]^{2D}\cdot[1-m/T]^{\lceil\tau\rceil}
\\
&\qquad\leq\label{EQ:PROB-D}
[e\cdot T/D]^{2D}\cdot\exp(-\lceil\tau\rceil\cdot m/T).
\end{align}
We note that \cref{EQ:PROB-A} maintains the requirement that $f$ and $h$ share the same classification of the first $\lceil\tau\rceil$ subjects and disagree more than $m$ times during the run phase,
\cref{EQ:PROB-A} to \cref{EQ:PROB-B} is the product of an upper bound on the number of possible pairs of functions and an upper bound on the probability that a random permutation produces at least $m$ errors, 
\cref{EQ:PROB-B} to \cref{EQ:PROB-C} makes use of Sauer's Lemma, and relies on the properties of binomial coefficients,\footnote{%
Specifically, $(T-m)/T\geq (T-m-z)/(T-z)$.}
and \cref{EQ:PROB-C} to \cref{EQ:PROB-D} makes use of the exponential bound, $1+z\leq\exp(z)$.
Since \cref{EQ:PROB-D} does not depend upon the sequence in which subjects arrive, the same bound applies for the unconditional probability over all possible sequences.
By setting
\begin{equation}
\label{EQ:M LEVEL}
m
=
\frac{T}{\lceil\tau\rceil}\cdot[2D\cdot\ln(e\cdot T/D) + \ln(3/\delta)]
\end{equation}
we determine that the error accumulated by using $\mathfrak{G}$ in place of $\mathfrak{F}$ is bounded from above by $m$ with probability at least $1-\delta/3$.

Combining Eqs. (\ref{EQ:RT_decomposition}), (\ref{EQ:Chernoff}), (\ref{EQ:Term_iv_bound}), and (\ref{EQ:M LEVEL}), we obtain that for a universal constant $c_2 > 0$,
\begin{align}
R_{T}/M
&\leq\label{EQ:SECOND PART-A}
\lceil\tau\rceil+c_2\sqrt{[T-\lceil\tau\rceil]\cdot[2D\cdot\ln(\lceil\tau\rceil\cdot e/D) + \ln(3/\delta)]}+m
\\
&\leq\label{EQ:SECOND PART-B}
\lceil\tau\rceil+c_2\sqrt{T\cdot[2D\cdot\ln(T\cdot e/D)+\ln(3/\delta)]+T\cdot\ln(3/\delta)}+m\\
&\leq\label{EQ:SECOND PART-C}
\lceil\tau\rceil+c_2\sqrt{m\cdot\left\lceil\tau\right\rceil+T\cdot\ln(3/\delta)}+m
\end{align}
with probability at least $1-\delta$.
We can then minimise the right-hand side of \cref{EQ:SECOND PART-C} by setting $\tau$ such that $m=\tau^{2}/\lceil\tau\rceil$.
The statement of \cref{THM:EXP4.P-VC} then follows. 
\end{proof}

\begin{proof}[Proof of \cref{THM:EXP4.P-LES}]
We follow the proof of \cref{THM:EXP4.P-VC}, modifying those parts that make reference to $\mathfrak{F}$. 
Specifically, we exploit the assumption that $\mathfrak{F}=\mathfrak{L}_{J}$ to obtain a tighter bound on its cardinality by invoking $\textrm{Harding}(t,J)$ rather than Sauer's lemma. We omit the details for brevity.

\end{proof}
\end{appendix}

\bibliographystyle{ecta}  
\bibliography{bibliography}

\end{document}